\documentclass[final,5p,times,twocolumn]{elsarticle}

\usepackage{amssymb}
\usepackage{algorithmic}
\usepackage[ruled,vlined]{algorithm2e}
\usepackage{setspace}
\usepackage{amsmath}
\usepackage{mathtools}
\newtheorem{theorem}{Theorem}[section]
\newtheorem{lemma}[theorem]{Lemma}
\newcommand{\R}{\mathbb{R}}

\usepackage{soul}
\usepackage{color}
\usepackage{xcolor}
\usepackage{booktabs}
\usepackage{hyperref}
\hypersetup{pdfauthor=author}

\usepackage{tikz}
\usepackage{pgfplots}
\pgfplotsset{compat=1.17}
\usetikzlibrary{decorations.pathreplacing}

\usepackage{caption}
\usepackage{subcaption}

\usepackage[mode=buildnew]{standalone}

\SetCommentSty{mycommfont}

\begin{document}

\begin{frontmatter}



\title{Federated Neural Collaborative Filtering}


\author[inst1,inst2]{Vasileios Perifanis\corref{cor1}}
\ead{vperifan@ee.duth.gr}
\author[inst1,inst2]{Pavlos S. Efraimidis}
\ead{pefraimi@ee.duth.gr}

\affiliation[inst1]{organization={Department of Electrical and Computer Engineering, Democritus University of Thrace},
            addressline={Kimmeria}, 
            city={Xanthi},
            postcode={67100}, 
            country={Greece}}

\affiliation[inst2]{organization={Institute for Language and Speech Processing, Athena Research Center},
            addressline={Kimmeria}, 
            city={Xanthi},
            postcode={67100}, 
            country={Greece}}
\cortext[cor1]{Corresponding author}

\begin{abstract}
In this work, we present a federated version of the state-of-the-art Neural Collaborative Filtering (NCF) approach for item recommendations. The system, named FedNCF, enables learning without requiring users to disclose or transmit their raw data. Data localization preserves data privacy and complies with regulations such as the GDPR. Although federated learning enables model training without local data dissemination, the transmission of raw clients' updates raises additional privacy issues. To address this challenge, we incorporate a privacy-preserving aggregation method that satisfies the security requirements against an honest but curious entity. We argue theoretically and experimentally that existing aggregation algorithms are inconsistent with latent factor model updates. We propose an enhancement by decomposing the aggregation step into matrix factorization and neural network-based averaging. Experimental validation shows that FedNCF achieves comparable recommendation quality to the original NCF system, while our proposed aggregation leads to faster convergence compared to existing methods. We investigate the effectiveness of the federated recommender system and evaluate the privacy-preserving mechanism in terms of computational cost.
\end{abstract}




\end{frontmatter}


\section{Introduction}
Recommender systems have been widely used for creating personalized predictive models that help individuals identify content of interest \cite{resnick1997recom}. Such systems collect various features of an individual, such as demographic characteristics, ratings to items (explicit feedback), or user interactions with items (implicit feedback) \cite{bobadilla2013recom}. Their goal is to provide future preferences based on past interactions and they are widely adopted in various domains such as e-commerce and online streaming services. The most straightforward technique for recommendations generation is collaborative filtering (CF) \cite{mnih2008mf, rendle2012bayesian}.

In the case of centralized learning, a service provider should collect user profiles along with their past interactions in a datacenter. Transferring users' private data is not easy to accomplish due to the restrictions imposed by regulations and laws, such as the GDPR \cite{yang2019fl}. To overcome the problem of storing such information in a centralized server and enhance users' privacy, Google has proposed a new technique called federated learning (FL) \cite{mcmahan2017fl}. In this machine learning setting, the computation is transferred on the user's side and after local training, weight updates are outsourced to the coordination server for aggregation. Since the federated setting facilitates machine learning without requiring the transmission of users' data, there are great promises for both privacy and fast distributed computations.

FL has attracted considerable interest in both industry and academia. However, techniques such as matrix factorization (MF) and neural networks, in the context of federated recommenders, remain highly unexplored. Moreover, private information can still be leaked in FL. More precisely, after local computations, clients should transmit their computed parameters to the coordination server for aggregation. The most popular aggregation method is the Federated Averaging (FedAvg) algorithm \cite{mcmahan2017fl}, in which clients transmit their updates in plain format. In horizontal federated MF, users share the same latent feature space and thus, private interactions can be inferred by an adversary \cite{Gao2020PTMF}. The coordination server can observe the differences in item updates between the clients' transmitted parameters and the previous global model to infer their observations \cite{Melis2019FeatureLeakage}. Hence, significant privacy concerns remain, which may prevent the system from reaching its goals. 

Motivated by the advances in FL and users' privacy concerns, this work explores the application of CF in a FL setting. Recent work on recommender systems has shown the feasibility of applying deep neural architectures to improve the quality of recommendations. One of the most successful systems is the NCF \cite{he2017neuralcollab} approach, which combines MF with deep neural architectures. In this work, we extend the state-of-the-art NCF method in a federated environment. In short, federated CF allows participants in the learning process to compute the weights of their model locally. Then, instead of the raw profile data, only the calculated weights are transmitted to the central entity, which aggregates the received parameters to generate the global model for the next round.
\begin{table*}[hbt!]
  \centering
  \begin{tabular}{ llc }
    \textbf{Algorithm}     &   \textbf{Description} & \textbf{This Work}\\
    \midrule
    FedAvg \cite{mcmahan2017fl}   &   Federated Averaging   &     \\
    SecAvg \cite{Bonawitz2017SecureAggregation}   &  Secure Aggregation   &     \\
    MF-FedAvg   &   FedAvg adapted to Matrix Factorization   &   \checkmark   \\
    MF-SecAvg  &   SecAvg adapted to Matrix Factorization  &   \checkmark\\
    SimpleAvg   &   Non-weighted Aggregation  &   \\
    \bottomrule
  \end{tabular}
  \caption{Notation and description for the aggregation algorithms used in this work.}
  \label{tab:aggregation_abbr}
\end{table*}

Federated aggregation algorithms such as FedAvg, were designed to enable the coordination server to perform an averaging on client updates that reflect the global objective. However, these algorithms are mainly suitable for neural networks, where users update all the model weights. In contrast, in CF tasks, users only update the item vectors observed in their local dataset. Hence, simply integrating an aggregation algorithm in CF will lead to inconsistent updates, resulting in decreased convergence or/and low quality. To overcome this limitation, in this paper, we consider an extension to the aggregation process of FedAvg, called MF-FedAvg, to handle the updates of the latent factors architecture. We experimentally show that the proposed variant leads to faster convergence and higher recommendation quality.

Except for the inconsistency in updates due to the integration of a federated aggregation algorithm in CF, many of the existing works, such as \cite{ammad2020MF, lin2020FedRec}, rely on the fact that users' data do not leave the local devices. However, MF-based models use an embedding layer to represent the item profile\footnote{We use the terms matrix and profile interchangeably, while the vector term, concerns a row in the matrix.}. A participant updates an item's vector in the matrix only when the corresponding item is observed in the local dataset. Therefore, the transmission of weight updates in plain format to the coordination server can reveal a user's preferences \cite{Melis2019FeatureLeakage, zhang2021survey}. To overcome this problem, we argue that integrating a Secure Multiparty Computation (SMC) aggregation based on a variant of FedAvg to latent factor models addresses privacy concerns compared to the transmission of raw weights. Another common privacy-preserving aggregation method is the utilization of homomorphic encryption schemes \cite{aono2017he}. These approaches, however, suffer from high computational complexity \cite{chai2020FMF}. A third approach to preserve the privacy of the participants is to prevent a passive entity from analyzing the received parameters by adding noise to achieve differential privacy guarantees \cite{Wei2020DiffPriv}. However, such privacy-preserving mechanisms come at a high utility cost. Hence, we use the Secure Aggregation (SecAvg) protocol \cite{Bonawitz2017SecureAggregation}, which is a SMC scheme, to enable a privacy-preserving aggregation. This scheme does not require heavy computation tasks on the user's side and provides equivalent quality to the direct transmission of model updates.

The aggregation of the transmitted updates is a crucial step in the FL pipeline, as it generates the new global parameters and is strongly related to the privacy of the participants \cite{zhang2021survey}. The algorithms used in the rest of this paper for achieving weights' aggregation and their description are summarized in Table \ref{tab:aggregation_abbr}. The learning algorithms adapted to the federated setting are marked with the prefix `Fed'.

We summarize our main contributions as follows:
\begin{itemize}
    \item We provide an extension to the FedAvg algorithm \cite{mcmahan2017fl} to handle the latent parameters of MF. While FedAvg has demonstrated its success both in the literature and in real-world applications in the context of neural networks, e.g., \cite{Bonawitz2019FLSystem}, we experimentally show that it leads to quality degradation in embedding-based models such as MF.
    \item The original FedAvg requires each participant to transmit the calculated updates in plain format. Therefore, FedAvg may sacrifice privacy for utility. We argue that integrating the SecAvg protocol \cite{Bonawitz2017SecureAggregation} as a privacy-preserving mechanism addresses privacy concerns against honest but curious (HBC) entities without sacrificing the recommendation quality. We also discuss the SMC aggregation scheme with respect to the k-anonymity requirement \cite{Samarati1998kanon, Samarati2001kanon, Sweeney2002kanon} per training round in the context of federated CF.
    \item We adapt the NCF approach to the FL setting for next item predictions. Unlike other solutions that have evaluated a federated CF setting, FedNCF leverages the non-linearity of neural networks to improve recommendations. To the best of our knowledge, this is the first work that analyzes the NCF system for the federated setting.
    \item We study the effectiveness of our approach on four real-world datasets and compare the FedNCF with the centralized NCF to validate its recommendation quality.
\end{itemize}
Our results show that FedNCF is a viable approach as it achieves comparable recommendation quality to the centralized NCF while improving users' privacy without requiring high communication and computational overhead.

\paragraph{\textbf{Organization}}
The remainder of the paper is structured as follows: Section \ref{Related Work} describes the preliminaries, including an introduction to MF, a discussion on machine learning models of the NCF approach, the scope of FL and similar work in federated CF. Section \ref{Federated Neural Collaborative Filtering} introduces FedNCF, which is an adaptation of NCF \cite{he2017neuralcollab} to the FL setting and details the privacy-preserving aggregation algorithm. The recommendation quality of FedNCF and the computation cost of the privacy-preserving scheme are evaluated in Section \ref{Experiments}. Finally, Section \ref{Conclusion} summarizes and concludes our work.

\section{Related work}
\label{Related Work}
In this section, we discuss the preliminaries for matrix factorization and the algorithms included in the Neural Collaborative Filtering framework. We then introduce the concept of federated learning and summarize methods that enable privacy-preserving training on the user side. Finally, we provide an overview of related privacy-preserving collaborative filtering approaches.

\subsection{Matrix factorization}
The goal of CF algorithms is to suggest new items to users based on their past behavior. In a typical scenario, the service provider has access to a set of $M$ users, $U = \{u_1, u_2, ..., u_M\}$ and a set of $N$ items, $I = \{i_1, i_2, ..., i_N\}$. Each user $u_i$ has interacted with a subset of items $n$. The interaction generated by user $i$ on item $j$ is represented as $r_{ij} \in \R$. Similarly, a matrix $R \in \R^{M \times N}$ represents the user-item interactions \cite{sarwar2001itemcollab}. The objective of a CF system is to provide a ranked list of top-$K$ items that a given user has not interacted with and that are suitable according to the user's preferences.

One of the most effective CF algorithms is MF and is based on latent factors \cite{bell2007Netflix}. The user-interaction matrix $R$ is decomposed to $X$ $\in \R^D$ and $Y \in \R^D$ matrices, where $D$ denotes the dimension of the latent space. In MF, the similarity of two users can be measured by the inner product of the matrices $X$ and $Y$ \cite{koren2009mf} and thus, $R \approx X \times Y$.

\subsection{Neural collaborative filtering}
The MF model estimates an interaction $r_{ij}$ as the inner product of the latent vectors. He et al. \cite{he2017neuralcollab}, however, argued that the product vector is inefficient in formulating users' similarity and showed that this limitation can be overcome by learning the interaction function using deep neural networks. First, they presented a generalized MF (GMF) model that uses embedding layers to obtain the latent user-item vectors. Then, the latent vectors are fed into a linear layer, which outputs the predicted score using the sigmoid function. Their second model is a multilayer perceptron (MLP) with at least one hidden layer. In this architecture, the user and item latent vectors are concatenated into a single vector and the output is then fed to the hidden layers. Finally, they showed that the fusion of the linear GMF model and the non-linear MLP model, namely, NeuMF, can lead to higher quality recommendations and faster convergence. In NeuMF, the GMF outputs the product of latent vectors and the MLP feeds the concatenation of the latent vectors into the deep neural network. The two outputs are concatenated in the last hidden layer, where a prediction is made.

\subsection{Federated learning}
Federated learning is a machine learning technique that allows the training of models in decentralized environments. The main idea behind this learning setting is that different entities can collaboratively train a model under the coordination of a central server without sharing their data. Unlike traditional machine learning, which requires users to transmit their data, FL enables a higher level of privacy, as the model is trained locally on each device. After the clients’ operations, weight updates are sent back to the central server without revealing the raw data \cite{kairouz2019AdvFL}. Upon receiving the updates, the coordination server performs an aggregation \cite{mcmahan2017fl} to achieve the learning objective.  

\subsubsection{Federated averaging}
The most popular technique for weights aggregation is the FedAvg algorithm \cite{mcmahan2017fl}. Briefly, after some local gradient descent iterations, participants transmit their local updates along with the number of local training instances to the aggregator. The aggregator updates the global parameters by:
\begin{equation}
\label{fedavg}
    w_{t+1} \gets \sum_{i=1}^{|c|}\dfrac{n_i}{n}w^i_{t+1},
\end{equation}
where $|c|$ is the number of selected participants in a training round, $n_i$ is the number of local training instances of a participant, $n=\sum_{i=1}^{|c|}n_i$ is the total number of training instances and $w_{t+1}^i$ is the local updates generated by the participant $i$.

\subsubsection{Privacy-preserving federated learning}
Although FedAvg does not require high computational and communication costs \cite{mcmahan2017fl, Bonawitz2019FLSystem}, simply transmitting weight updates cannot ensure the privacy of the participants. In this context, privacy-preserving techniques include SMC, homomorphic encryption and differential privacy approaches \cite{Truong2021PPF}. In the former, clients exchange some random values to hide their updated parameters from external entities. In homomorphic encryption, the coordination server performs aggregation over encrypted data, while clients can operate on the encrypted global model. Finally, another approach is to add noise to the calculated parameters to hide the presence of a training instance.

In this work, we integrate the SecAvg protocol \cite{Bonawitz2017SecureAggregation}, which is a SMC approach that only allows the disclosure of the sum of the weights updates while the intermediate results from each participant remain secret. Unlike homomorphic encryption approaches, such techniques do not require high costs on the coordination server and the user side and different from differential privacy, they preserve the model's utility. Hence, we use the SecAvg scheme, which is secure against HBC entities and can also handle user dropouts, a property that is essential in FL since network failures can be a common event \cite{Bonawitz2019FLSystem}. In learning environments based on embedding layers such as MF, SecAvg can effectively prevent the coordination server from inferring the user's observations. The coordination server can only learn that one or more participants have interacted with a particular item, but it cannot deduce any information about a participant's identity.

\subsection{Privacy-preserving recommenders}
In this section, we summarize the proposed privacy-preserving CF methods in both centralized and federated settings.

One of the earliest works on privacy-preserving MF was presented in \cite{Nikolaenko2013PPMF}. More precisely, a SMC technique based on garbled circuits was proposed to enable a MF model generation without requiring users to reveal their interactions. An extension of that protocol was presented in \cite{Kim2016EfficientPPMF}, in which a fully homomorphic encryption construction proposed to build the privacy-preserving recommender. Both works rely on a two-party setting, where users encrypt their data and the execution is performed without revealing user interactions. In the former protocol, the execution time and the communication cost between the two external parties may be prohibitive, requiring about 1.5 hours of execution time for 4096 tuples of interactions and almost 40 GB per iteration. In the latter work, the setting remained the same in terms of the entities involved in the computation, with a strong decrease in execution time and communication cost. However, both works require users to transmit their data (in encrypted form) and the model's generation is based on a centralized approach, which differs from our approach. 

In the context of centralized learning, Berlioz et al. \cite{Berlioz2015DPMF} applied noising techniques to achieve differential privacy guarantees \cite{dwork2013DP}. They showed that perturbing the inputs approximates the quality of the pure computation, while they found that deferentially private stochastic gradient descent (DP-SGD) with tight privacy guarantees leads to increased quality loss. Closer to distributed learning, \cite{Shin2018PEMF} proposed a randomized algorithm that transforms users' ratings with a probability while decoupling the user and item vectors on the user and server sides, respectively. The communication cost is further reduced by allowing users to share only a gradient update of the item profile. However, the learning process requires the transmission of all users' gradients in a single iteration, which is impractical in a real-world scenario. Unlike previous centralized or distributed approaches, strict differential privacy guarantees with high-quality model generation are hard to achieve in the federated setting \cite{Wei2020DiffPriv} and need further investigation. Preliminary results on FedNCF by applying the DP-SGD algorithm \cite{Abadi2016DPSGD} to obtain differential privacy guarantees, before weights aggregation, showed high quality degradation and slow convergence. Therefore, the integration and formalization of differential privacy in FedNCF needs further research and we leave it for future work.

In the context of FL, Ammad et al. \cite{ammad2020MF} proposed a federated CF algorithm based on implicit feedback. Although users keep their data locally, they send raw gradient information to the coordination server, which can lead to information leakage as proved by \cite{chai2020FMF}. An extension of the federated CF approach for explicit feedback is presented in \cite{lin2020FedRec}. Chai et al. \cite{chai2020FMF} incorporated a homomorphic encryption scheme to enhance the privacy of the participants. However, homomorphic encryption introduces significant overhead as users need to decrypt the received item profile, perform local updates, encrypt the updated item profile and transmit it to the aggregator. 

In each of the above approaches, the coordination server performs the weight update procedure immediately, when one or more updates are received \cite{ammad2020MF, lin2020FedRec, chai2020FMF}, running an asynchronous version of Stochastic Gradient Descent (SGD) \cite{Dean2012AsyncSGD}. As a result, these systems do not require an aggregation algorithm such as FedAvg, while an optimizer can internally perform the update procedure. Therefore, these approaches are asynchronous FL frameworks \cite{Chen2019OnlineFL}, where the coordination server waits for the updates from one to several clients, which can result in staleness \cite{Damaskinos2020Fleet}, i.e., the received updates may be computed on an outdated model.

More recently, Yang et al. \cite{yang2021FCMF} proposed FCMF, in which two entities who hold ratings from users, jointly train a model under HE. While their system protects user privacy, this approach is not applicable in our scenario, since in this paper, the federated setting is intended to prevent the transmission of users' data to an external entity. Hence, we study the application of CF in a horizontal federated setting \cite{zhang2021survey}. We focus on the aggregation step, where the coordination server averages the updates received from the participants. Previous work on privacy-preserving horizontal federated CF \cite{chai2020FMF}, emphasized on the prediction task by updating the item profile when an update is received. Although this technique is promising in FL, the staleness problem can lead to the prevention of convergence \cite{Jiang2017Staleness}. Besides, HE approaches incur heavy computational and communication costs \cite{chai2020FMF}, which may prevent the system from reaching its goals.

FedNCF mitigates the staleness constraint by leveraging synchronous FL: in the current global step, the coordination server chooses a subset of available clients and transmits the current global model; the clients receive the global parameters and perform computations on the latest model's release. Moreover, we employ an efficient SMC scheme to achieve a privacy-preserving aggregation, which does not require heavy computational overhead compared to homomorphic encryption approaches and does not introduce quality degradation that is inevitable when using noising techniques.

Following the security definition for horizontal FL systems \cite{yang2019fl}, in this work, we assume that the participants and the coordination server are non-colluding HBC entities, i.e., they follow the protocol faithfully but try to infer additional information \cite{goldreich2009HBC}.

\section{Federated neural collaborative filtering}
\label{Federated Neural Collaborative Filtering}
In this section, we introduce the FedNCF system and the MF-SecAvg approach, which is an integral part of FedNCF. We discuss the level of privacy achieved and show that the k-anonymity requirement is satisfied in our approach.

\subsection{Problem definition}
We consider a scenario where multiple users ($M > 2$), each holding a private dataset, want to jointly build a horizontal federated recommender without revealing their raw data. The observations are only collected locally and never transmitted. In addition, the community a user belongs to could heavily influence the corresponding interactions. Hence, the distributed nature of FL poses several statistical challenges \cite{zhao2018noniid}, as users may hold an arbitrary number of local observations that can vary in distribution. In the remainder of this paper, we assume that the data each participant owns are independent and non identically distributed (non-iid). This assumption is natural in FL since each participant can have an arbitrary number of training instances, while the local datasets may not be representative of the overall distribution. We plan to investigate the non-iid characteristic in future work further, e.g., using a clustered federated learning approach \cite{Sattler2021Clustered} to group participants according to their distributions.

 \begin{table*}[htb!]
  \centering
  \begin{tabular}{ llc }
    \textbf{Notation}     &   \textbf{Description}\\
    \midrule
    $M$ &   Number of users\\
    $N$ &   Number of items\\
    $D$ &   Latent dimension\\
    $u \in \{1, 2, ..., M\}$    &   User's id\\
    $i \in \{1, 2, ..., N\}$    &   Item's id\\
    $r_{ij} \in \{0, 1\}$  &   User's $i$ interaction with item $j$\\   
    $I$ &   Shared items latent vectors\\
    $U_i$ &   User's latent vector\\
    $N$ &   Neural network's weights\\
    $\mathcal{P}_i = \{U_i, I, R_i\}$   &   User's set of preferences\\
    $C \subseteq \mathcal{P}$ &   Number of available clients \\
    $c \subseteq C$ & Randomly selected clients\\
    $MI$    & Masked calculated weights for the item profile\\
    $MN$    & Masked calculated weights for the neural network\\
    $MP$    & Masked calculated vector of observations\\
    $E$     & Local epochs\\
    \bottomrule
  \end{tabular}
  \caption{Notations and description for the parameters in FedNCF.}
  \label{tab:notations}
\end{table*}
The primary goal of such a system is to generate a top-N recommendation list for each user based on local computations without violating participants' privacy. The personal data required for the federated training process can vary and are based on the learning objective. For instance, a service such as Google Maps would collect a user's visits according to the geographical distance with a venue in the local devices. The visits could be collected without interrupting the user's actions. Due to privacy concerns, the raw data are not transmitted to a server, but end users can jointly utilize the local observations for generating a high-quality recommendation model using a FL approach. The notations for the parameters in the FedNCF system are given in Table \ref{tab:notations}.

\newdefinition{definition}{Definition}
\begin{definition}
Given $M$ participants, where each participant represents an individual user, $\mathcal{P}_{u \in \{1 , ..., M\}} = \{u, i, r\}$, where $u \in \{1, ..., M\}$ is the user id, $i \in \{1, ..., N\}$ is the item id and $r \in \{0, 1\}$ is a binary value, federated CF tries to generate a recommendation model by incorporating users' past preferences on a shared item profile $I$, while minimizing the disclosure of interacted items for each participant.
\end{definition}

In the FedNCF solution, user interactions never leave the local devices, while the raw computed weights are masked with a SMC scheme after a local update.

In our problem, the observed interactions $r$ are binarized data indicating whether a user interacted with an item or not. Implicit feedback can indirectly reflect a user's preferences and is, therefore, easier to collect. In MF, a user's latent vector $U$ is essential in the inference stage. However, the transmission of users' latent vectors raises privacy concerns \cite{ammad2020MF}. The coordination server can directly infer private interactions after accessing the transmitted updates. Following previous work \cite{ammad2020MF, yang2021FCMF}, we ask each participant to maintain the corresponding $U_i$ locally.

\subsection{The FedNCF framework}
\label{The FedNCF framework}
\begin{figure*}[htb!]
    \centering
    \includegraphics[width=0.7\textwidth]{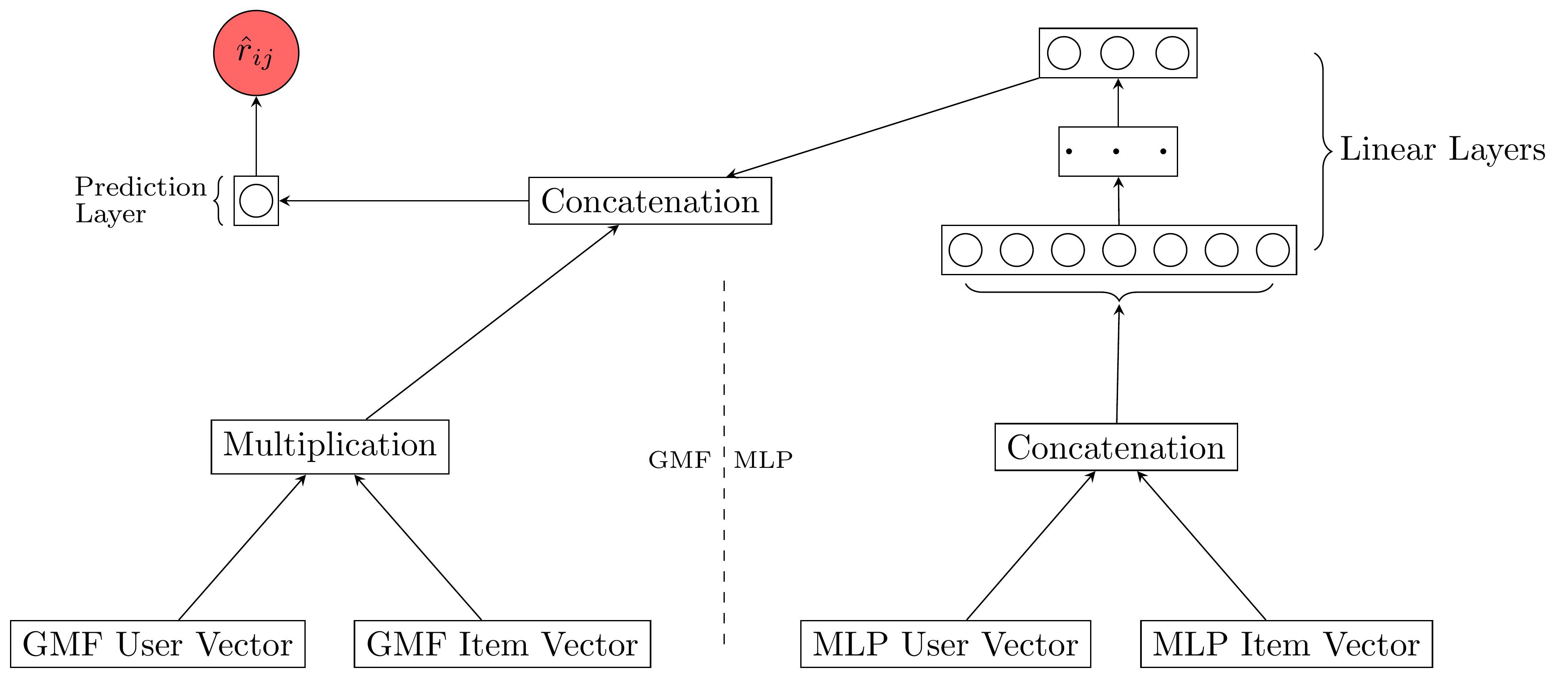}
    \caption{Architecture overview of the NeuMF model.}
    \label{Arch_overview}
\end{figure*}
The FedNCF framework comprises three FL algorithms: a generalized MF (FedGMF), a multilayer perceptron (FedMLP) and a fusion of FedGMF and FedMLP (FedNeuMF). These models are the federated adaptations of the corresponding NCF approach \cite{he2017neuralcollab}. Fig. \ref{Arch_overview} presents the architecture of the NeuMF model. In the input layer, a user $i$ and an item $j$ are represented as vectors, i.e., embeddings. The NeuMF model contains the GMF and the MLP model and thus, the input layer consists of two user embeddings and two item embeddings, which are updated independently in the training stage. The first part, i.e., the GMF model, performs element-wise multiplication between the two embeddings. The other part, i.e., the MLP model, concatenates the embeddings and the resulting output is passed to linear layers. The outputs of GMF and MLP are then concatenated into a single vector, which serves as the final input to the prediction layer. At the end of the computation, a prediction $\hat{r}_{ij}$ is generated, indicating the preference of user $i$ to the item $j$. Discarding the MLP model and connecting the multiplication output to the prediction layer, we end up with the GMF model. Similarly, discarding the GMF model and connecting the final output of the linear layers to the prediction layer, we get the MLP model. 

In FedNCF, a user is an agent that holds private interactions and can train a machine learning algorithm. The coordination server maintains some global parameters and controls the learning process. Within the server, there are two components: the user selection and the aggregation functions. The user selector is responsible for selecting a set of clients to perform a local update and the aggregator is responsible for averaging their updates to form the new global parameters. The communication between the participants and the coordination server is performed over a secure channel (SSL/TLS). In this work, our primary focus is to evaluate the recommendation quality of the federated system against the centralized one. Hence, we assume that all participants are available at any point of time.

At time step $t$, the coordination server initiates a \textit{training plan}. The plan describes the task that includes the model's hyper-parameters, e.g., the mini-batch size $B$. Then, the model's current global parameters $W_t$ are prepared for transmission. Note that in the first round, $W_0$ is generated at random. These parameters are sent to randomly selected $c \subseteq C \subseteq \mathcal{P}$ clients, where $C$ denotes the available clients in the current time step. 
\begin{figure*}[!htb]
    \centering
    \includegraphics[width=0.7\textwidth]{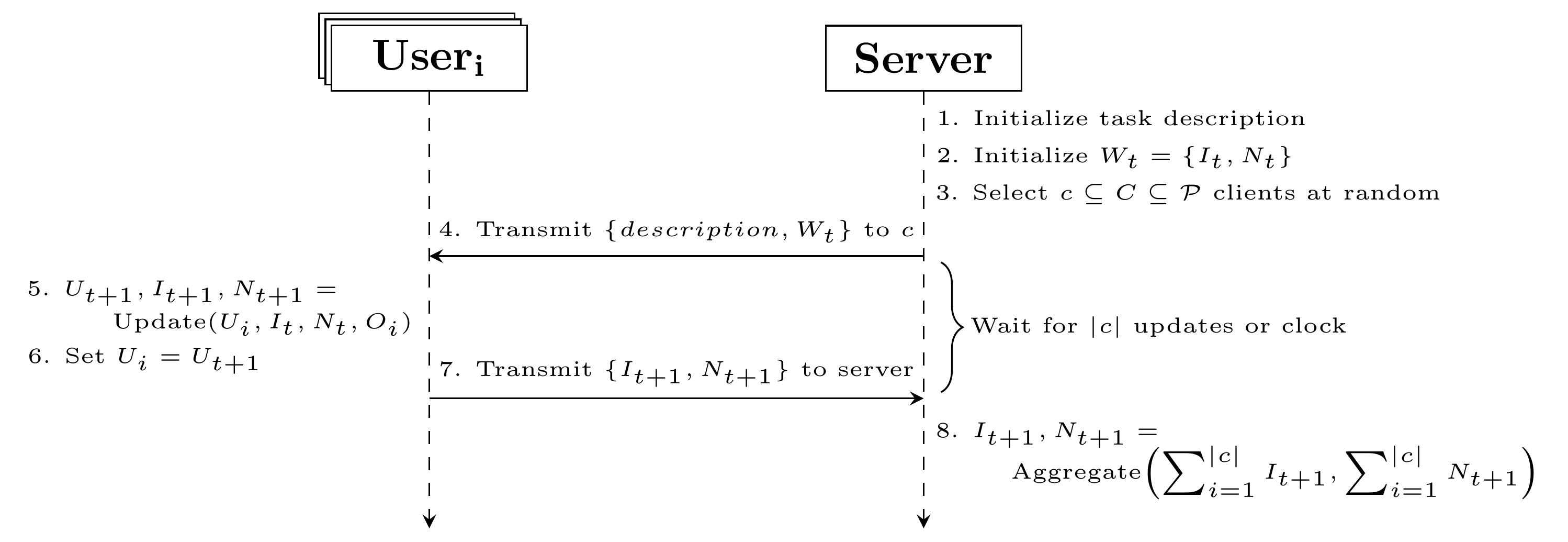}
    \caption{Order of operations and parameters exchange between the coordination server and clients in an aggregation round.}
    \label{communication_model}
\end{figure*}

The selected clients download the training plan and use their local data along with some local steps, referred to as \textit{local epochs}, to update the global parameters. In our models, there are three different types of weights on each client side: (1) the user profile $U_i$, (2) the item profile $I_i$ and (3) the neural network's weights $N_i$. The transmission of the parameters $I_i$ and $N_i$ to the coordination server is sufficient for the learning objective. Since the user profile raises privacy concerns, it is never outsourced. Hence, the updated item profile and the neural network's weights are transmitted to the coordination server for aggregation after a local update. 

The coordination server waits for the transmission of the updated weights and performs an aggregation step after receiving $|c|$ updates or after a period of time has elapsed. Finally, it generates the new global parameters and the training process is repeated until model convergence. Fig. \ref{communication_model} shows the interactions between the coordination server and the selected clients as well as the computation steps performed on each side. 

We refer to the process of aggregating $|c|$ updates as an \textit{aggregation round}. In centralized learning, an epoch (or a training round) has passed when all available data are fed into the learning algorithm. To fairly compare the federated with the centralized setting, we refer to the pass over all participants as \textit{global round}.

\subsection{Weights aggregation}
\label{Weights Aggregation}
The FedAvg algorithm \cite{mcmahan2017fl} is the most popular method for weights aggregation, which has shown its effectiveness in the context of neural networks and tasks such as classification \cite{Bonawitz2019FLSystem}. However, the aggregation algorithm must handle the additional latent vectors in a MF task. Unlike neural networks, where a user influences every part of the network weights, in most cases, the local updates in MF only affect a small part of the item profile. Utilizing the original FedAvg algorithm to handle the updates of the embedding layers of MF will lead to aggregated parameters being relatively close to the parameters from the previous global model. This is due to the specificity of the updates in the item profile. For instance, if a user has not interacted with an item, the vector of that item is not affected by the local update. The FedAvg algorithm includes this parameter in the calculation process and hence, the update for this item remains close to the previous value and as a result, convergence is slowed down. 

Let us consider a scenario with two clients $P_1$ and $P_2$, where $P_1$ has interacted with 150 items and $P_2$ with 170 items. For simplicity, the dimension of the latent factors is $D=1$. We focus on a single item $i_1$ with an original weight of 0.04. The participant $P_1$ who interacted with this item, updates the corresponding weight to 0.047, while $P_2$ retains the original weight, i.e., $i_1$ is not observed in the local dataset. The FedAvg averaging calculation (eq. \ref{fedavg}) generates an aggregated result of 0.0433. Similarly, a simple aggregation results in an aggregated output of 0.0435. However, neither averaging strategy reflects the actual update of $i_1$. Since only $P_1$ has interacted with this item, the aggregated result should only account for the update of this client. Based on this observation, we argue that aggregating the item profile weights by considering the number of users who actually updated an item will lead to faster convergence and higher recommendation quality. 

Except for the inconsistency in the updates of embedding-based models, the original FedAvg algorithm is vulnerable to an HBC coordination server because participants transmit their calculated values in plain format \cite{chai2020FMF}. A plausible way to achieve a privacy-preserving aggregation without compromising quality is to adopt an efficient secure aggregation scheme such as SecAvg \cite{Bonawitz2017SecureAggregation}. At a higher level, SecAvg enables the coordination server to blindly compute the sum of the participants' updates, without revealing the generated weights that correspond to each individual. The protocol operates by pairing each user with every other selected participant in an aggregation round. Then, each pair agrees on a random seed and after the agreement, each user performs a simple calculation based on their rank in the user order \cite{Bonawitz2017SecureAggregation}. In SecAvg, a total order is assumed, i.e., an identifier is assigned to each participant and based on these identifiers, users perform addition or subtraction on their calculated weights with random parameters generated from the agreed seed. However, simply integrating this protocol in the case of MF for a weighted aggregation will lead to the inconsistency in updates as described earlier. 

In the next section, we introduce MF-SecAvg, which handles the update procedure of the three federated models by decomposing the aggregation into a MF-based step and a neural network-based step. In summary, the MF-based aggregation concerns the averaging of the item profile. The protocol operates independently on each item by considering the number of users who have updated this item. Regarding the weights of the neural architecture, our approach scales to the weighted aggregation of FedAvg \cite{mcmahan2017fl}. Discarding the neural architecture, MF-SecAvg operates on a traditional MF algorithm. The entire process is integrated with the SecAvg protocol \cite{Bonawitz2017SecureAggregation} to minimize privacy concerns against HBC entities. 

\subsection{Secure aggregation in matrix factorization}
\paragraph{\textbf{Item Profile Update}}
The first concern is the update procedure of the item profile, which is updated so that a participant contributes only to the components that correspond to their interactions. Hence, the profile can be updated on the server side by averaging participants' weight updates that include an item in the local training process.

In particular,  at time step $t$, each selected user $i \in c$ generates the weight updates that correspond to the item profile $I_{t+1}^i$, using some local gradient descent iterations. Before transmitting their updates to the coordination server, each user agrees on random seeds with every other selected participant in the aggregation round. Then, the agreed seeds are used to generate a random matrix $IR_{ij}$, according to the size of the item profile, where $j \in c, j \ne i$. Note that a user $i$ generates a random matrix for every other participating client. Finally, each user masks the updated item profile by:
\begin{equation}
   \label{secavg_clients}
    MI_{t+1}^i = I_{t+1}^i + \sum_{i \in c:i<j}IR_{ij} - \sum_{i \in c: i>j}IR_{ji},
\end{equation}
where $IR_{ij}$ is the generated random matrix using the agreed $seed_{ij}$ between user $i$ and user $j$ in an ordered pair of users $(i, j), i < j$ and $MI_{t+1}$ is the masked calculated weights. The coordination server, after collecting each $MI_{t+1}^c$ computes the following:
\begin{equation}
    \label{secavg_server}
    \begin{aligned}
    I_{t+1}^{sum} &= \sum_{i \in c} MI_{t+1}^i\\
            &= \sum_{i \in c}I_{t+1}^i + \sum_{i \in c:i<j}IR_{ij} - \sum_{i \in c: i>j}IR_{ji} \\
            &= \sum_{i \in c}I_{t+1}^i.
      \end{aligned}            
\end{equation}
The generated $I_{t+1}^{sum}$ parameter contains the sum of the weight updates that corresponds to the item profile, which needs to be aggregated. In the simplest form of an aggregation step, the server can generate the aggregated weights as follows:
\begin{equation*}
    I_{t+1} = \frac{MI_{t+1}^{sum}}{|c|},
\end{equation*}
where $|c|$ is the number of selected participants at time step $t$. However, this kind of aggregation is not adjusted for each item independently and can lead to slower convergence as mentioned earlier.

Inevitably, the coordination server should be aware of the number of participants who interacted with specific items to perform the averaging step for each item's vector according to the number of users who updated this item. A naive solution for clients is to disclose their observed interactions. However, this approach would lead to information leakage since the coordination server can observe users' interactions and infer additional information. To overcome this limitation, the procedure of the SecAvg protocol can be further exploited to enable users to mask a vector that contains their preferences. 

In particular, the participating clients generate a vector $P$ containing ones on the indices that correspond to the items updated by the local training process. Then, similar to the generation of random matrices for the item profile, each user generates a random vector $PR_{ij}$ using $seed_{ij}$. Finally, users mask their interactions according to eq. \ref{secavg_clients} and transmit the masked vector $MP_{t+1}$ to the coordination server. Subsequently, the coordination server can correctly calculate the number of participants that interacted with an item without inferring the identity of a participant. The masks will be canceled when added together (eq. \ref{secavg_server}), while the participants' plain training instances are maintained secret. 

\paragraph{\textbf{Neural Architecture's Update}}
\begin{figure*}[htb!]
    \centering
    \includegraphics[width=0.7\textwidth]{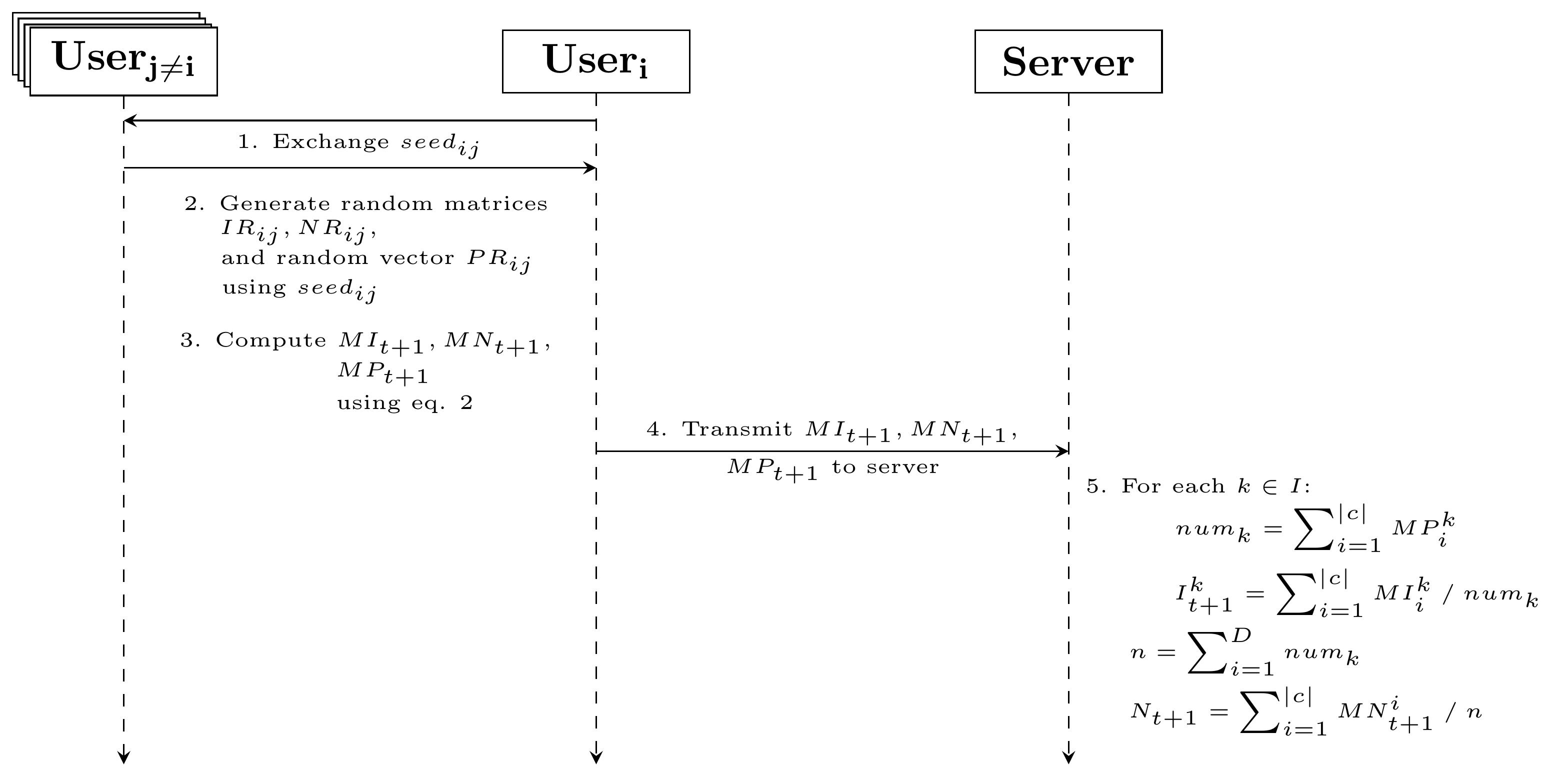}
    \caption{Overview of MF-SecAvg.}
    \label{MF-SecAvg_overview}
\end{figure*}
As mentioned earlier, FedNCF contains a neural architecture for each of the three learning algorithms. Applying the MF-SecAvg approach to the neural network's weights is a simpler task than handling the item profile. After a local training process, each user multiplies the updated weights by the number of local training instances and performs a computation based on the SecAvg protocol. First, the local $N_{t+1}^i$ is multiplied by $n_i$, where $n_i$ denotes the number of local training instances. Then, each participating user generates random weights using the agreed $seed_{ij}$. Finally, they transform the local neural network weights according to eq. \ref{secavg_clients} and the masked weights $MN_{t+1}$ are transmitted to the coordination server.

The coordination server receives the masked $MN_{t+1}$ parameter from each participating client, generates the sum of the masked weights and divides the resulting output by the total number of training instances $n$. The total number of training instances is known to the coordination server since it can be calculated from the received $MP_{t+1}$ vectors. The division operation between the sum of the masked neural network weights $\sum_{i=1}^{|c|}MN_{t+1}^i$ and the total number of training instances $n$, is equivalent to the FedAvg procedure (eq. \ref{fedavg}).

The secure aggregation process of MF-FedAvg is summarized in Fig. \ref{MF-SecAvg_overview}. After local training, each pair of users that participated in the current aggregation round exchanges a seed variable. Then, each user generates the random $IR_{ij}$ and $NR_{ij}$ matrices as well as a random vector $PR_{ij}$, which correspond to random parameters for the item profile, neural network weights and observed instances in the local dataset, respectively, using the agreed $seed_{ij}$. Note that a user $i$ generates $2 (|c| - 1)$ random matrices and $(|c| - 1)$ random vectors, according to the number of users that participated in the current training round. The generated random parameters are utilized to transform the plain updates $I_{t+1}$ and $N_{t+1}$ as well as the local vector of training observations $P$ according to eq. \ref{secavg_clients}. The masked parameters $MI_{t+1}, MN_{t+1}$ and $MP_{t+1}$ are transmitted to the coordination server, who performs the MF-based and neural network-based aggregations. The operations on the masked parameters hide a participant's local instances and result in an equivalent aggregated output to the plain transmission of the parameters. Therefore, unlike homomorphic encryption, SMC approaches similar to SecAvg do not require heavy computation and communication costs since simple operations are introduced to blur the plain updates of the participating clients. Besides, unlike noising techniques, the model's utility is maintained compared to the plain transmission of the updates.

\paragraph{\textbf{Computation and communication cost}}
\begin{figure*}[htb!]
    \centering
    \includegraphics[width=0.7\textwidth]{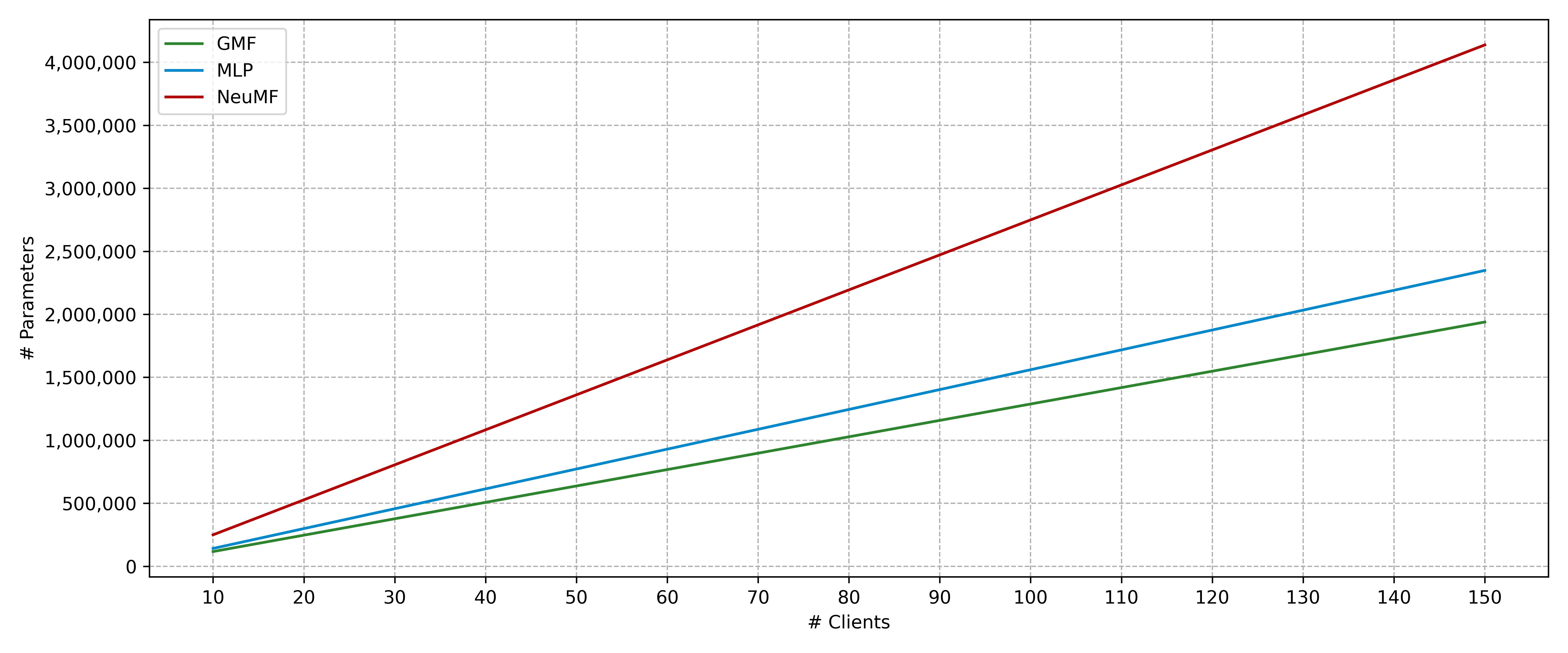}
\caption{Parameters growth on a single user by the number of participating clients $|c|$ with MF-SecAvg. The dimension $D$ is fixed to 12 and the number of items $|I|$ to 1000.}
\label{number_generated_vs_param}
\end{figure*}
After local training and seeds exchange (step 1 in Fig. \ref{MF-SecAvg_overview}), users locally generate $2 (|c| - 1)$ random matrices and $|c| - 1$ random vectors, where $|c|$ denotes the number of users who participated in the current round.  In each considered model, the number of parameters grows with the specified dimension size $D$ and the number of items $|I|$ in the profile. For the GMF model, which contains a single linear layer (with a single processing unit), each user generates 
\begin{equation*}
    \left(|c| - 1\right) \cdot \left(D \cdot |I| + D + 1 + |I|\right)
\end{equation*}
parameters based on the agreed seeds, where $D \cdot |I|$ represents the number of values in the item profile, $D + 1$ the number of inputs and biases on the neural architecture and $|I|$ the number of parameters for the random vector of interactions. Discarding the neural architecture, the model falls into traditional MF, while the number of parameters that are being generated shrinks to $\left(|c| - 1\right) \cdot \left(D \cdot |I| + |I|\right)$. The MLP model contains an architecture with at least one hidden layer. Hence, the number of parameters to be generated grows with specified hidden layers and the number of processing units in each hidden layer. More precisely, each user generates
\begin{equation*}
    \begin{multlined}
    \left(|c| - 1\right) \cdot \left(D \cdot |I| + 2D \cdot h1 + \left(\sum_{i=1}^{n-1} h_i \cdot h_{i+1}\right) + h_n + \right.\\
    \left. \left(\sum_{i=1}^n h_i\right) + 1 + |I|\right)
    \end{multlined}
\end{equation*}
parameters, where $2D \cdot h_1$ is the input size, $\sum_{i=1}^{n-1} h_i \cdot h_{i+1}$ are the number of weights and $\sum_{i=1}^n h_i + 1$ are the biases of the network, with $h_i$ denoting the number of processing units on the $i$th hidden layer. As the NeuMF model is the concatenation of GMF and MLP, $2D \cdot |I|$ parameters are required for the item profile. Hence, the total parameters to be generated after the seeds agreement are:
\begin{equation*}
    \begin{multlined}
    \left(|c| - 1\right) \cdot \left(2D \cdot |I| + 2D \cdot h1 + \left(\sum_{i=1}^{n-1} h_i \cdot h_{i+1}\right) + D + h_n + \right.\\
    \left. \left(\sum_{i=1}^n h_i\right) + 1 + |I|\right).
    \end{multlined}
\end{equation*}

The number of parameters to be generated with MF-SecAvg, roughly grows linearly with respect to one of the parameters that contribute to the computation while keeping the rest of the parameters constant. In the training process of a federated recommender, the dimension size $D$, the number of items $|I|$ and the (possible) number of processing units $h_i$ are agreed upon before the initialization of the computation. Consequently, the growth of the parameters concerns the number of clients that participate in an aggregation round. Fig. \ref{number_generated_vs_param} shows the increase on the parameters on single user using the MF-SecAvg approach. The number of clients $|c|$ is from $\{10, 11, ..., 100\}$, the dimension size and the number of items are fixed to 12 and 1000, respectively and the size of hidden layers for the MLP and NeuMF models are fixed to $h = \{48, 24, 12, 6\}$. The corresponding number of parameters using a traditional MF is almost equal to the GMF model and therefore, it is omitted from Fig. \ref{number_generated_vs_param}. Even in settings with a great number of clients, the local devices need some additional storage, e.g., considering $|c|=150$, the additional storage requirement is roughly 40 MB. Note that the growth only concerns the second step (Fig. \ref{MF-SecAvg_overview}) in the process of MF-SecAvg, while the parameters are being removed after the execution of the protocol. In addition, a linear execution of the protocol, i.e., by generating the random parameters for one participating user at a time and executing the third step of the protocol, prevents the additional storage requirements on the user side.

As a final step under MF-SecAvg, clients perform simple matrix operations and distribute the final output. These operations concern an element-wise addition or subtraction of the computed weights after a local update with the randomly generated parameters based on their identified order. Hence, the total number of operations for executing the third step of the protocol is equivalent to the number of parameters generated at the second step. In the experimental evaluation, we show that the execution of the second and third step of the protocol introduces negligible computational cost. Regarding the communication cost after the execution of the protocol, the final output is equivalent to the plain transmission of the local computed weights in terms of kilobytes (KB). For instance, in the GMF model, the transmitted parameters from each client are $D \cdot |I| + D + |I| \approx 200 KB$ with $D=12$ and $|I| = 1000$, thus avoiding heavy communication costs compared to homomorphic encryption approaches \cite{chai2020FMF}. 

Putting it all together, the main steps of the FedNCF learning process are as follows:
\begin{enumerate}
    \item In the first round, the coordination server randomly initializes the weights corresponding to the item profile, $I_0$, and the neural architecture, $N_0$. In later rounds, it prepares the aggregated parameters computed in the previous global step for the next round. Then, it randomly selects $c \subseteq C$ online participants to perform a local update and informs them about the training plan, i.e. the model's parameters.
    \item The participants download the current global parameters and perform a specified number of local epochs $E$.
    \item Each pair of participants agrees on a random seed. Then, each user generates $|c| - 1$ random matrices for the item profile and neural network weights and random vectors according to the agreed seeds.
    \item Each participant performs a calculation to the computed weights according to MF-SecAvg, which is an extension of the SecAvg \cite{Bonawitz2017SecureAggregation} protocol to handle the item profile updates.
    \item The coordination server waits for the transmission of the masked parameters within a specified time and aggregates the transmitted updates by decomposing the aggregation into MF-based and neural network-based steps.
\end{enumerate}
\begin{algorithm}[!hbt]
\DontPrintSemicolon
  \caption{Federated collaborative filtering using MF-SecAvg on coordination server.}
  \label{alg:mfsecavg}
  \textbf{Server executes:\\}
    initialize $N_0, I_0$\\
    \For{each round t=1,2,...}{
        \tcp{Select $c \subseteq C \subseteq \mathcal{P}$ clients at random}
        $c \gets$ \text{set of $C$ online clients} \\
        \For{each client $i \in c$} {
            \tcp{Get masked parameters $MN, MI, MP$ from each participant}
            $MN_{t+1}^i, MI_{t+1}^i, MP_{t+1}^i$ $\gets$ \text{LocalUpdate($N_t$, $I_t$, $c$)}\\
        }
        Initialize empty vector $instances[]$\\
        \For{each $k \in I$} {
            \tcp{Get the number of participants who updated each item}
            $num_k \gets \sum_{i=1}^{|c|} MP_i^k$\\
            \tcp{Aggregate masked item's updates}
            $I_{t+1}^k \gets \dfrac{1}{num_k} \: \sum_{i=1}^{|c|} MI_i^k$\\
            $instances[k] \gets num_k$
        }
        \tcp{Calculate the total number of training instances}
        $n \gets \sum_{k=1}^|I| instances[k]$\\
        \tcp{Aggregate masked neural network's updates}
        $N_{t+1}$ $\gets$ $\dfrac{1}{n}\sum_{i=1}^{|c|}MN_{t+1}^i$\\
    }
\end{algorithm}
We summarize the learning process for a CF task using MF-SecAvg, which is the privacy-preserving integration in FedNCF, in Algorithms \ref{alg:mfsecavg} and \ref{alg:mfsecavg_client}. Algorithm \ref{alg:mfsecavg} shows the operations on the server side, while Algorithm \ref{secavg_clients} shows the training procedure and the generation of the masked parameters on the user side.

\begin{algorithm}[!hbt]
\DontPrintSemicolon
  \caption{Federated collaborative filtering using MF-SecAvg on user; $O$ denotes the local observations.}
  \label{alg:mfsecavg_client}
    \textbf{Client $\mathbf{i}$ executes:\\}
        Generate $P$ that contains ones for the observed instances in $O$\\
        $\mathcal{B}$ $\gets$ \text{split $O$ into batches of size $B$}\\
        \For{each local epoch e=1,2,...,E}{
            \tcp{Update the item profile and neural network weights}
            \For{batch $b \in \mathcal{B}$}{
                $I \gets I - \eta \nabla \mathcal{L} (I;b)$\\
                $N \gets N - \eta \nabla \mathcal{L} (N;b)$
            }
        $MI \gets I, MN \gets N, MP \gets P$\\
        Exchange seed with participant $j \in c, j \neq i$\\
        \For{each $j \in c$}{
            Generate $IR_{ij}, NR_{ij}, PR_{ij}$ at random using $seed_{ij}$\\
            \If{$i < j$}{
                $MI \gets MI + IR_{ij}$\\
                $MN \gets MN + NR_{ij}$\\
                $MP \gets MP + PR_{ij}$
            }
            \Else{
                $MI \gets MI - IR_{ij}$\\
                $MN \gets MN - NR_{ij}$\\
                $MP \gets MP - PR_{ij}$
            }
        }
        \KwRet $MI, MN, MP$ to server
  }
\end{algorithm}

\subsection{Security analysis}
The required parameters during a local computation in FedNCF are the local interactions $r_{ij}$, the local user vector $U_i$, the current global item profile $I_t$ and the global neural network's weights $N_t$. At the end of a local computation, each selected participant $i \in c$ transmits the masked local parameters $MI$ and $MN$ along with the vector $MP$ to the coordination server for aggregation. Recall that in FedNCF, participants keep their corresponding user vectors locally. 

Below, we define the privacy level achieved in the FedNCF solution with the integration of MF-SecAvg. We consider two potential adversaries against the FedNCF system: (1) the coordination server and (2) the participants $\mathcal{P}_{u \in \{1,...,M\}}$.

\begin{itemize}
    \item \textbf{Steps 1-2.} In the first two steps, the coordination server transmits the aggregated global parameters and the participants perform computations based on their local observations. The aggregated parameters hide the weight updates of a particular user while the user profile is maintained in the local devices. Therefore, there is no information leakage.
    \item \textbf{Steps 3-4.} Before transmitting the local updates, participants agree on random seeds in steps 3 and 4 to generate random parameters and mask their updates. Following the security analysis in \cite{Bonawitz2017SecureAggregation}, the masked parameters computed independently by each participant look random if the agreed seed between users ($i, j$), $seed_{ij}$ is generated uniformly at random. Each participant under MF-SecAvg performs addition or subtraction to the calculated local updates with the randomly generated parameters using the agreed $seed_{ij}$, based on the identified order (eq. \ref{secavg_clients}). The final output is transmitted for aggregation. Therefore, the users in the current training round are only aware of their agreed random seeds and subsequently for the generated random parameters, their output after a local training operation and the global model's weights. Thus, they cannot learn anything other than their own updates.
    \item \textbf{Step 5.} In step 5, the masked parameters are transmitted to the coordination server for aggregation. The random matrices hide the original weights calculated by each individual in a way that the sum of the masked weights is equal to the sum of the plain weight updates generated by the participants (equation \ref{secavg_server}). Hence, the coordination server learns only the masked weights of a participant, the sum of the transmitted parameters and the number of users who updated a particular item's vector in the profile. Although the number of participants who interacted with certain items is known, the coordination server cannot learn the identities of users who interacted with these items.
\end{itemize}

In summary, the masked parameters hide the users' updates and therefore, the learning process under the MF-SecAvg protocol does not reveal any information about the preferences of the selected participants in an aggregation round. After generating the aggregated parameters, the coordination server transmits the model to $c$ randomly selected participants. The parameters of an aggregated model do not pose an immediate threat since the noise introduced by aggregating the weights from the previous round hides the presence of a single user's interactions.

\subsubsection{k-Anonymity of FedNCF}
The concept of k-anonymity is introduced by Samarati and Sweeney \cite{Samarati1998kanon, Samarati2001kanon, Sweeney2002kanon}. Briefly, a database release satisfies k-anonymity if every record is indistinguishable from at least $k-1$ other records, ensuring that individuals cannot be identified by linking attacks.

In the context of FedNCF, k-anonymity is achieved by ensuring that no fewer than $k=c$ participants can be associated with a given item's vector update.

We distinguish three entities in the FedNCF system: (i) the participants in an aggregation round, whose privacy needs to be protected, (ii) the coordination server and (iii) an HBC adversary. Note that an HBC adversary can be either a participant or the coordination server.

The participants control both their local datasets and local training iterations. In FedNCF, participant interactions are associated with the item's profile updates. We assume that in each aggregation round, the adversary if fully aware of the global parameters and attempts to perform a linking attack, i.e., the goal is to link a particular user to an item update.

\begin{lemma}
FedNCF enhanced with MF-SecAvg ensures the k-anonymity of the participants against an HBC adversary in an aggregation round.
\end{lemma}
\newproof{pf}{Proof}
\begin{pf}
Recall that MF-SecAvg enables a blind calculation of the sum of the participants'  updates and the number of participants who updated a given item. The participants in each aggregation round only have access to the current global parameters and their local interactions. Even if the same group of users is considered in successive aggregation rounds, an HBC participant cannot associate another user's interactions with an item's update. Therefore, k-anonymity holds against HBC participants.

The coordination server collects and aggregates the masked parameters received from the selected users. The participants' weight updates are masked under the MF-SecAvg protocol in the collection phase. In the aggregation phase, the coordination server averages the sum of the masked updates for an item based on the number of participants that updated this item. Although the coordination server is aware of the number of participants who updated an item, it cannot infer which user(s) actually interacted with this item. Hence, k-anonymity holds against an HBC coordination server.
\end{pf}

\subsubsection{An Attack against the k-anonymity of FedNCF}
A limitation of FedNCF under the MF-SecAvg scheme is that an HBC coordination server can manipulate the random selection process to choose the same group of participants, except for one user, to identify this user's interacted items. For instance, at time step $t$, the coordination server learns that five of $|c|$ randomly selected users have updated an item's $i$ vector. In step $t+1$, it selects the same group of $c$ clients, except for one user. In the current round, it learns that four clients have updated this $i$'s vector. Hence, the coordination server can deduce a particular user's preference for a specific item and breaks the concept of k-anonymity. We argue that this limitation can be mitigated by the blurring property of CF algorithms with synthetic negative feedback.

\paragraph{\textbf{Blurring property of FedNCF}}
In implicit feedback CF, only positive interactions are provided. A common strategy to simulate negative interactions is to randomly sample some non-interacted items to augment a participant's training dataset. Negative sampling can also be extended to an explicit feedback scenario, e.g., by assigning a pseudo-rating to some randomly selected non-interacted items \cite{lin2020FedRec}. Therefore, even if an HBC coordination server can force the selection of the same group of participants, negative sampling blurs a user's actual interactions. The blurring property hides the actual preferences of users. Therefore, the coordination server cannot presume that a specific item is included in the observed interactions of a particular user.

Although a negative sampling strategy can enhance participants' privacy, an individual's behavior in a multi-round scenario can still be deduced while selecting the same group of participants. The interacted items are included in a participant's training set and hence, the observed item vectors are updated in each round, regardless of the negative interactions. In any case, the coordination server should select the same group of participants for multiple rounds to observe the group's behavior and then exclude the target participant for several consecutive rounds to determine which items are actually observed locally.

The complexity for such an attack is relatively high and may not be feasible in a real-world scenario since clients are not always available. Besides, this limitation can be overcome by transferring the selection process to another entity, which performs a random selection using a secure pseudo-random generator such as \cite{Varizani1984RandomGen}. Another option is to ask the participants to perform a group check operation by noticing the users' identities before the seeds agreement step under the MF-SecAvg scheme. A different approach may be to add noise in the local training step or to transmit the updated weights with a probability to provide differential privacy guarantees \cite{dwork2013DP}. However, noising techniques affect the model's quality and we leave integrating the concept of differential privacy in FedNCF for future work.

\begin{table*}[htb!]
  \centering
  \begin{tabular}{ lcccc }
    \textbf{Dataset}    &   \textbf{\#Interaction} & \textbf{\#Item} & \textbf{\#User}   & \textbf{Sparsity}  \\
    \midrule
    \textbf{MovieLens 100K}   &   100,000   &   1,682    &   943 & 93.7\%\\
    \textbf{MovieLens 1M}   &  1,000,209   &   3,706     & 6,040    & 95.53\%\\
    \textbf{Lastfm 2K}   &   185,650   &   12,454     &   1,600 & 99.07\%\\
    \textbf{Foursquare NY}   &   91,023   &   38,333     &   1,083 & 99.78\%\\
    \bottomrule
  \end{tabular}
  \caption{Evaluation datasets statistics.}
  \label{tab:datasetstatistics}
\end{table*}
\begin{table*}[htb!]
  \centering
  \begin{tabular}{ lccccc }
    \textbf{Dataset}    &   \textbf{\#Min} & \textbf{\#Max} & \textbf{Avg.}   & \textbf{St. dev.}   & \textbf{Variance}  \\
    \midrule
    \textbf{MovieLens 100K}   &   20   &   737    &   106.04 & 100.88   & 10,176\\
    \textbf{MovieLens 1M}  &  20   &   2,314     & 165.6    & 192.73   & 37,145\\
    \textbf{Lastfm 2K}   &   5   &   2,609     &   116.03 & 240.07  & 57,635\\
    \textbf{Foursquare NY}   &   9   &   714     &   84.05 & 47.83  &   2,287.83\\
    \bottomrule
  \end{tabular}
  \caption{Data quantity heterogeneity.}
  \label{tab:data_hetero}
\end{table*}
\section{Experiments}
In this section, we introduce the evaluation datasets and detail the experimental settings. Then, we assess the recommendation quality of the FedNCF system in terms of the utilized aggregation function and the distributed nature of federated learning. We then evaluate the three considered models concerning the trade-off between computational cost and recommendation quality and measure the impact of heterogeneity on training time. Finally, we conduct experiments by removing users to assess the robustness of FedNCF and evaluate the computational cost of the MF-SecAvg approach.
\label{Experiments}

\subsection{Evaluation settings}
We evaluate the FedNCF system on four real-world datasets in recommender systems: MovieLens 100K and
MovieLens 1M
\cite{Harper2015movielens},
Lastfm 2K
\cite{Cantador2011lastfm}
and Foursquare New York (NY)
\cite{Yang2015Foursquare}. These datasets are widely used in the literature to evaluate CF algorithms. The first two datasets are movie ratings; Lastfm 2K contains tagged artists and Foursquare NY contains users check-ins. We excluded users with less than 5 interactions in each dataset. Table \ref{tab:datasetstatistics} shows the characteristics of the four datasets. 

Each user in the federated setting may have interacted with an arbitrary number of items. To measure the data quantity heterogeneity, we report the minimum, maximum and average number of local observations as well as the standard deviation and the variance in the data quantity of the population on Table \ref{tab:data_hetero}. The range between the minimum and the maximum number of local instances and the high values of standard deviation and variance imply that the number of local observations is highly diverse. Subsequently, a quality decrease compared to the centralized setting and different training times on local devices due to data heterogeneity are expected.

For pre-processing, we follow a common practice in recommendation systems by converting numerical ratings into implicit feedback \cite{he2017neuralcollab, rendle2012bayesian, Hu2008implicit} for the MovieLens datasets. Lastfm 2K and Foursquare NY only contain user interactions and thus, they are already in an implicit format.

To evaluate the recommendation models, we adopt leave-one-out evaluation, a commonly used method in the literature \cite{rendle2012bayesian, he2017neuralcollab}. We hold out the last interaction for each user as validation data and utilize the remaining interactions for training. For a quick evaluation, we pair each ground truth item in the test set with 100 randomly sampled uninteracted items \cite{he2017neuralcollab}. Therefore, the CF task is transformed to rank the sampled negative items with the held-out item for each user.

The ranked list is evaluated using the \textit{Hit Ratio} (\textit{HR}) and \textit{Normalized Discounted Cumulative Gain} (\textit{NDCG}) metrics. Briefly, \textit{HR} calculates the occurrence of the ground truth item in the top-$K$ ranked items and \textit{NDCG} considers the position of the hit \cite{he2017neuralcollab, he2015trirank}. In this work, we report the \textit{HR} and \textit{NDCG} with $K=10$. In the evaluation, we compute both metrics for each user and report the average score.

\subsection{Implementation details}
\label{Implementation}
We implement our proposed method with PyTorch \cite{paszke2019torch}. The weight parameters in the neural architecture are initiated using Xavier initialization \cite{Glorot2010training}. For common hyper-parameters in all models, we set the latent dimension $D = 12$ and the hidden layers for the neural architectures to $h = \{48, 24, 12, 6\}$. In addition, all weights are learned by optimizing the BCE loss, where we sampled four negative instances per positive instance. For the optimization, we utilized the Adam optimizer \cite{kingma2014Adam} with a learning rate of $0.001$. In each experiment, we conduct 400 epochs of training for our models, with the local epochs fixed at $2$. Finally, grid search is applied to the FedGMF model to find the optimal $c$ from $\{10, 20, 50, 100, 200, 300,|P|\}$ participants per aggregation round, where $|P|$ is the total number of users present in the dataset. We found that selecting 20, 120, 50 and 50 participants for each dataset, respectively, leads to the highest quality models. The divergence in the model's quality when selecting a different constant $c$ can be attributed to the heterogeneity among participants, which leads to inconsistency among local models \cite{zhao2018noniid}.

\subsection{Aggregation function impact}
We first demonstrate the influence of the aggregation function, which is an integral step of the FedNCF approach. The primary purpose of this experiment is to illustrate the effectiveness of the weights aggregation procedure in a setting with additional learning parameters other than neural network weights. 
\begin{figure*}[htb!]
    \centering
    \begin{subfigure}[b]{0.49\textwidth}
        \centering
        \includegraphics[width=0.85\textwidth]{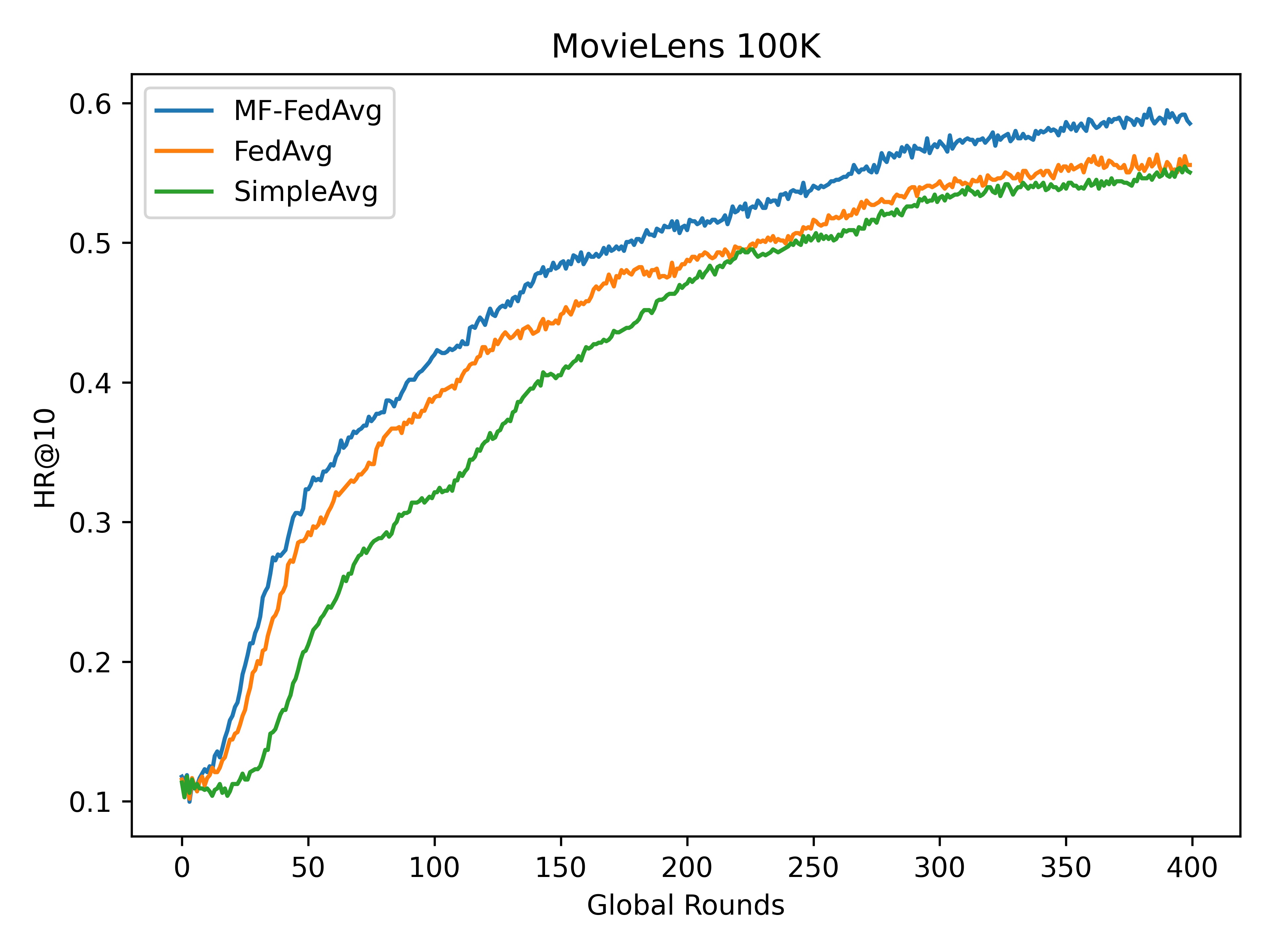}
    \end{subfigure}
    \begin{subfigure}[b]{0.49\textwidth}
        \centering
        \includegraphics[width=0.85\textwidth]{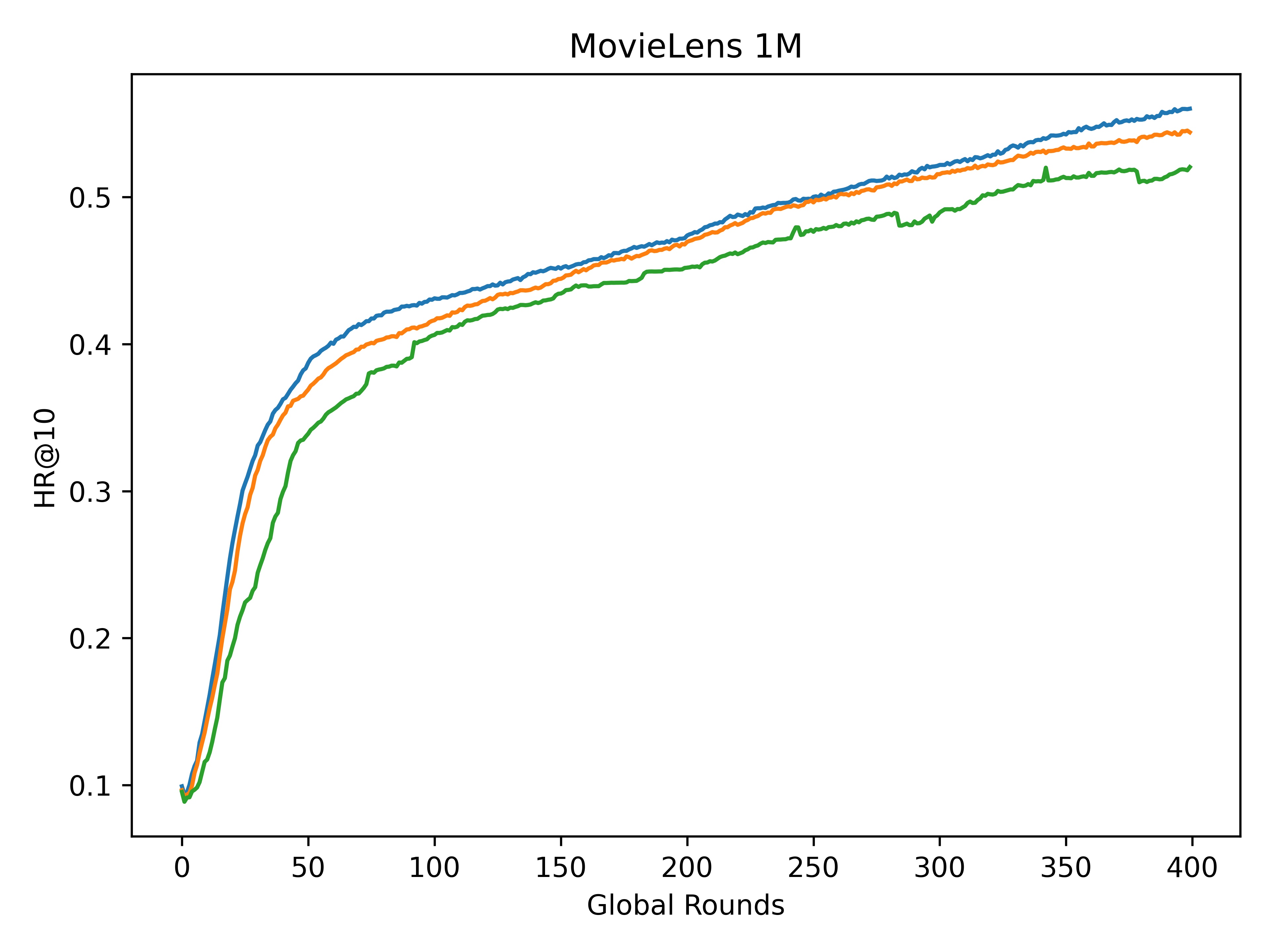}
    \end{subfigure}
    \begin{subfigure}[b]{0.49\textwidth}
        \centering
        \includegraphics[width=0.85\textwidth]{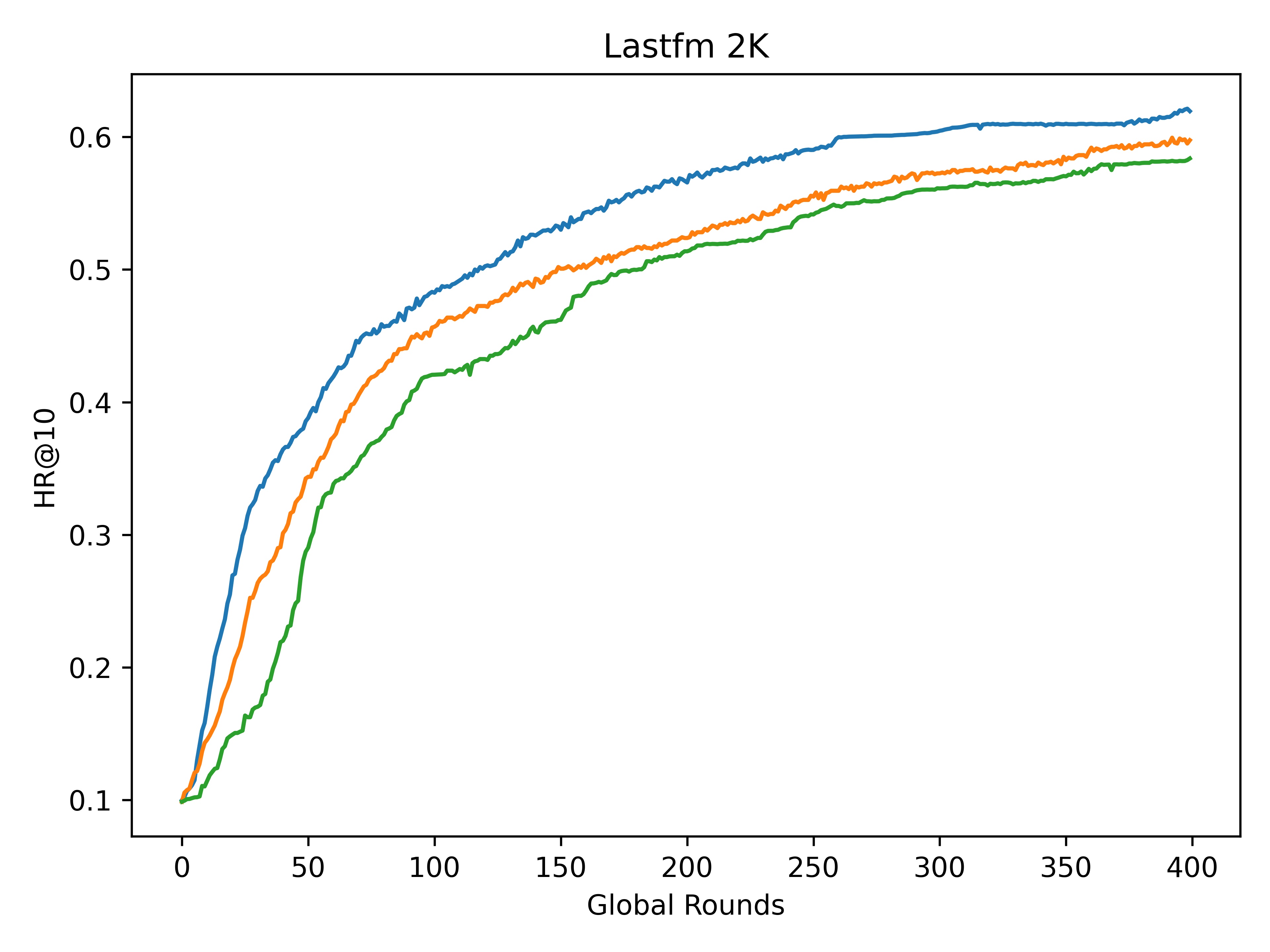}
    \end{subfigure}
    \begin{subfigure}[b]{0.49\textwidth}
        \centering
        \includegraphics[width=0.85\textwidth]{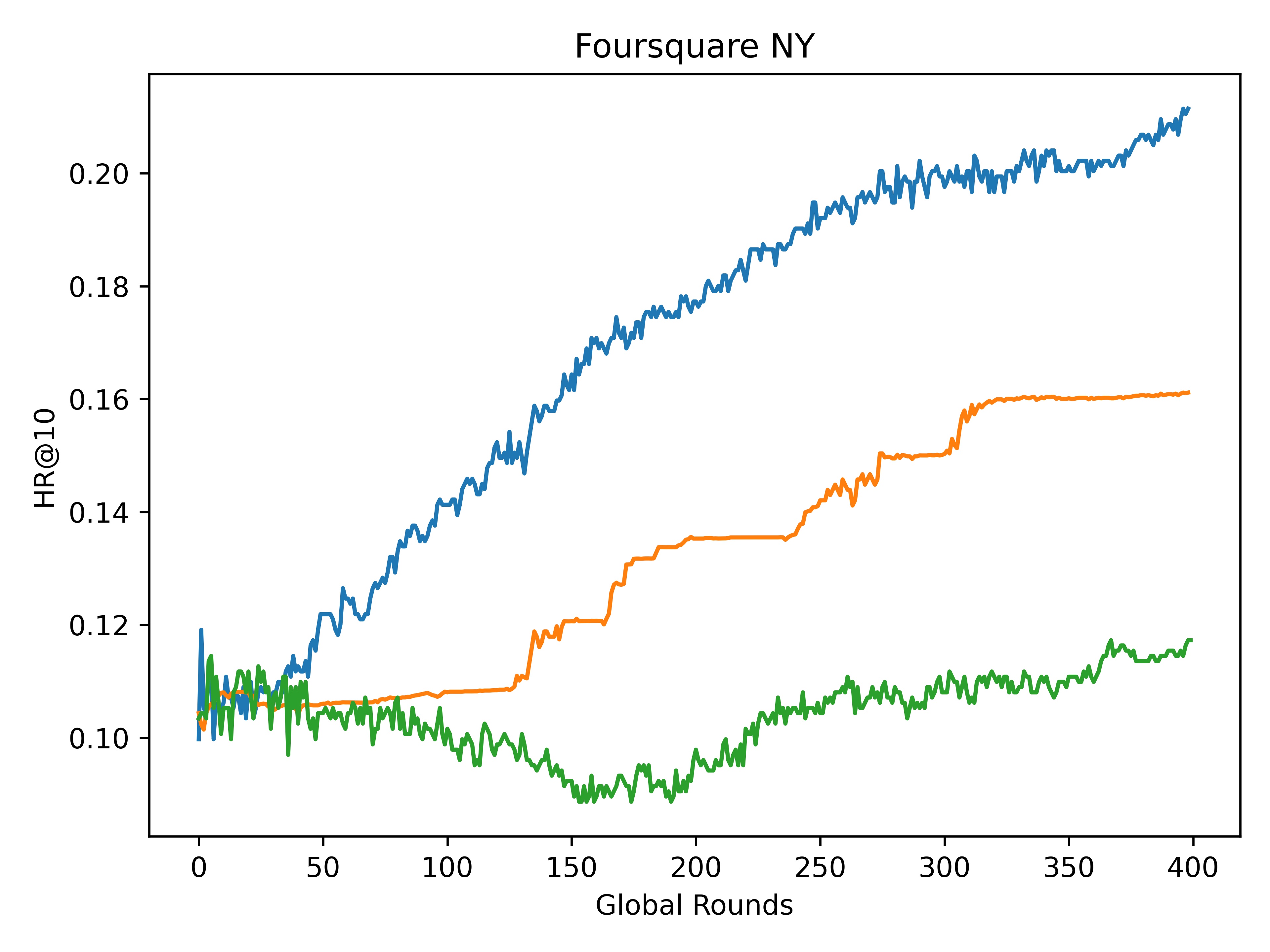}
    \end{subfigure}
\caption{Convergence speed of the MF-FedAvg, FedAvg and SimpleAvg aggregation methods.}
\label{Agg_impact_conv}
\end{figure*}

We compare the proposed aggregation method, which decomposes the averaging step into a MF-based and a neural network-based step, with the FedAvg and SimpleAvg algorithms, where SimpleAvg is a non-weighted averaging over the received parameters. Note that in this experimental evaluation, the aggregation is performed without integrating any privacy-preserving mechanism. Hence, our approach falls into plain weights transmission from clients to the coordination server, denoted as MF-FedAvg. Table \ref{tab:fed_agg_fun} shows the quality comparison between the three algorithms by reporting the average $HR@10$ over 5 experiments with different training and testing sets in the FedGMF model for each dataset using the same hyper-parameters.
\begin{table}[hbt!]
  \centering
  \begin{tabular}{ lccc }
    \textbf{Dataset}    &   \textbf{MF-FedAvg} & \textbf{FedAvg} & \textbf{SimpleAvg}  \\
    \midrule
    \textbf{MovieLens 100K}   &   \textbf{0.59}   &   0.56    &   0.55\\
    \textbf{MovieLens 1M}   &   \textbf{0.56}   &   0.54     &   {0.52}\\
    \textbf{Lastfm 2K}   &   \textbf{0.62}   &   0.59     &   0.58\\
    \textbf{Foursquare NY}  &   \textbf{0.21}   &   0.16     &   0.12\\
    \bottomrule
  \end{tabular}
  \caption{Influence on the recommendation quality using MF-FedAvg, FedAvg and SimpleAvg aggregation functions.}
  \label{tab:fed_agg_fun}
\end{table}

The recommendation quality shows that MF-FedAvg outperforms FedAvg, which outperforms SimpleAvg. Thus, there is evidence that the MF-FedAvg approach better fits the clients' updates in federated models based on MF than FedAvg, which is mainly applied to neural networks. 

MF-FedAvg also leads the model to faster convergence. Each of the three methods starts with an \textit{HR} close to $0.1$ in each of the four datasets, while MF-FedAvg converges faster than FedAvg. Fig. \ref{Agg_impact_conv} shows the convergence using the considered aggregation algorithms in the four datasets. At the end of the training iterations, we observe that the \textit{HR} quality of MF-FedAvg is 2-5\% higher than the corresponding \textit{HR} of the FedAvg method. The difference in the model's quality and the convergence speed can be attributed to the fact that the FedAvg algorithm heavily depends on the number of local training instances, which is inconsistent with item profile updates in MF. On the other hand, MF-FedAvg performs an aggregation to the item profile based on the number of users who updated each item, while the aggregation of the neural network weights is performed considering the weighted average of FedAvg. Therefore, the global parameters generation is consistent with the update process of MF with respect to the item profile and can lead the model to both higher quality and faster convergence.

\subsection{Data federation impact}
\begin{figure*}[htb!]
\centering
\includegraphics[width=0.85\textwidth]{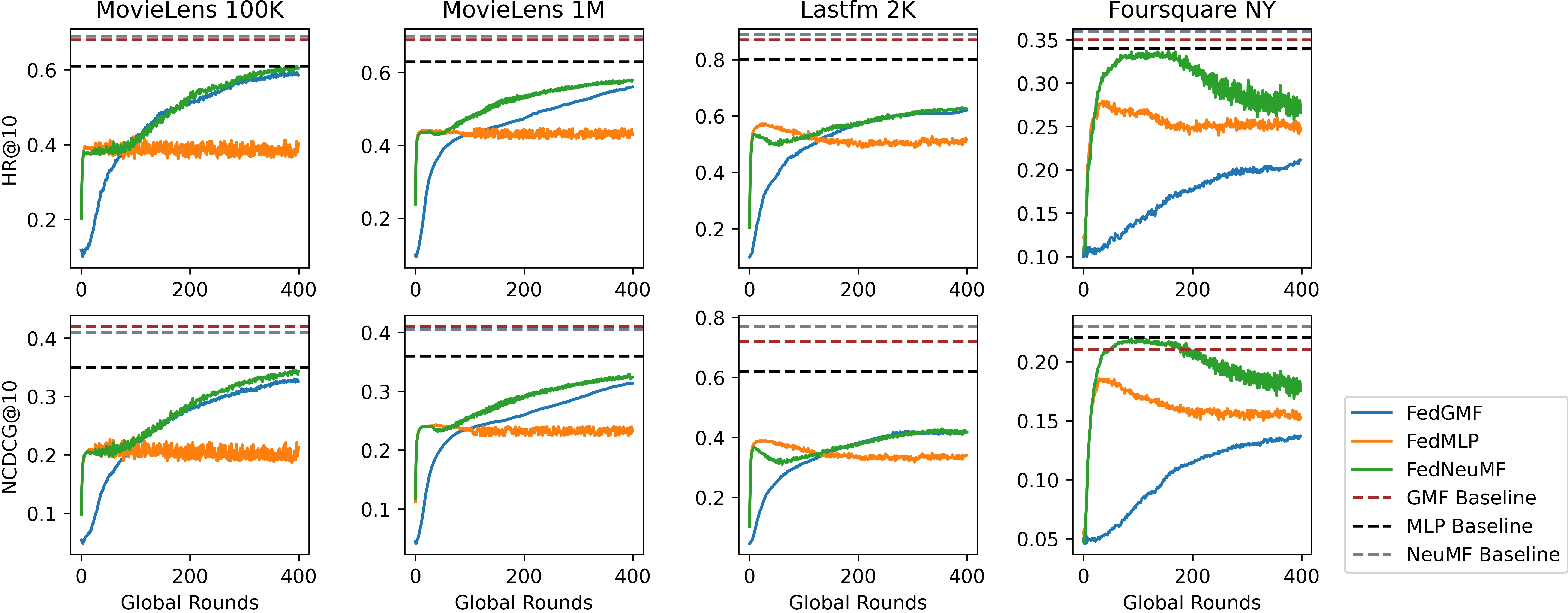}
\caption{Quality comparison and convergence speed of NCF and FedNCF.}
\label{NCFvsFedNCF}
\end{figure*}
Our primary goal in this experiment is to compare the recommendation quality and convergence speed of FedNCF with the centralized NCF system. We compare FedNCF with the centralized NCF by measuring the $HR@10$ and $NDCG@10$ metrics generated by each method. Fig. \ref{NCFvsFedNCF} shows the recommendation quality and the convergence speed of the considered models in both the centralized and federated settings after 400 epochs of training.

The recommendation quality of the centralized trained models is the upper bound for FedNCF. The highest \textit{HR} and \textit{NDCG} values in the NCF system are generated by the NeuMF model, while GMF provides almost equivalent recommendation quality. Similarly, FedNeuMF and FedGMF outperform FedMLP in the federated setting. In the Foursquare NY dataset, FedNeuMF and FedMLP overfit the training data, beginning from the 180th and 60th global round, while FedGMF has not reached its maximum quality after 400 global rounds. 

The FedMLP and FedNeuMF models improve convergence at the beginning of the federated training compared to the GMF model. However, FedNeuMF and FedGMF offer similar recommendation quality at the end of the training iterations in each dataset, while FedMLP fails to converge. This behavior of FedMLP can be attributed to the distributed nature of FL. In a centralized environment, the MLP model tries to capture the similarity between users given the user profiles to provide higher quality recommendations. However, in our setting, the user profiles are never transmitted and thus, the correlation cannot be sufficiently learned. In FedMLP, each user only owns its corresponding user vector and therefore, each training iteration leads to the inconsistency of the global learning objective, which prevents convergence. Hence, the NeuMF algorithm, which internally contains both the GMF and MLP models, exploits the fast convergence at the beginning of the training iterations of the MLP model and then utilizes the smaller steps of GMF to provide high quality recommendations. Although FedNeuMF provides better recommendations than FedGMF, in the next section, we argue that complex architectures are not preferable in a federated setting, as simple models such as MF can provide high quality recommendations without incurring heavy computational overhead.
\begin{table}[hbt!]
    \centering
    \begin{tabular}{ llcc }
        Dataset &   &   FedNCF  &   NCF\\
        \midrule
        &   &   \multicolumn{2}{c}{\textbf{GMF}}\\
        \textbf{MovieLens 100K} &   HR  &   0.59    &   0.68\\
        &   NDCG    &   0.33    &   0.42\\
        \textbf{MovieLens 1M}   &   HR  &   0.56    &   0.69\\
        & NDCG  &   0.31    &   0.41\\
        \textbf{Lastfm 2K}  &   HR  &   0.62    &   0.87\\
        & NDCG  &   0.47    &   0.72\\
        \textbf{Foursquare NY}  &   HR  &   0.21    &   0.35\\
        &   NDCG    &   0.14    &   0.21\\
        \midrule
        &   &   \multicolumn{2}{c}{\textbf{MLP}}\\
        \textbf{MovieLens 100K} &   HR  &   0.43    &   0.61\\
        &   NDCG    &   0.23    &   0.35\\
        \textbf{MovieLens 1M}   &   HR  &   0.44    &   0.63\\
        &   NDCG    &   0.24    &   0.36\\
        \textbf{Lastfm 2K}  &   HR  &   0.57    &   0.8\\
        &   NDCG    &   0.39    &   0.62\\
        \textbf{Foursquare NY}  &   HR  &   0.28    &   0.34\\
        &   NDCG    &   0.19    &   0.22\\
        \midrule
        &   &   \multicolumn{2}{c}{\textbf{NeuMF}}\\
        \textbf{MovieLens 100K} &   HR  &   0.61    &   0.69\\
        &   NDCG    &   0.34    &   0.42\\
        \textbf{MovieLens 1M}   & HR    &   0.58    &   0.7\\
        &   NDCG    &   0.33    &   0.41\\
        \textbf{Lastfm 2K}  &   HR  &   0.63    &   0.89\\
        &   NDCG    &   0.43    &   0.77\\
        \textbf{Foursquare NY}  &   HR  &   0.33    &   0.36\\
        &   NDCG    &   0.22    &   0.23\\
        \bottomrule
    \end{tabular}
    \caption{Quality comparison between FedNCF and NCF.}
    \label{tab:comparison_Fed_Cen}
\end{table}

The comparison of recommendation quality for the HR and NDCG metrics between FedNCF and NCF is given in Table \ref{tab:comparison_Fed_Cen}. The difference between FedNeuMF and NeuMF is 8\%, 12\%, 16\% and 3\% for the HR metric in each dataset, respectively. Although the difference between the two settings in the MovieLens 1M and Lastfm datasets is over 10\%, the federated models provide acceptable recommendation quality.

\subsection{Efficiency comparison}
\begin{table*}[htb!]
    \centering
    \begin{tabular}{ llcccc }
        & & &  & & \textbf{Comm.}\\
        \textbf{Dataset} & & & \textbf{FLOPs (K)} & & \textbf{Cost (KB)}\\
        \midrule
        & &  \textbf{Avg.} & \textbf{Min.} & \textbf{Max.}\\
        MovieLens 100K & GMF  & 17 & 3.2 & 117.9 & 21\\
        & MLP  & 1,510.4 & 285.6 & 10,524.4 & 33\\
        & NeuMF & 1,532.9 & 289.1 & 10,653.3 & 55\\
        \midrule
        MovieLens 1M & GMF  & 26.5 & 3.2 & 370 & 46\\
        & MLP  & 2,365 & 286 & 33,044 & 58 \\
        & NeuMF & 2,394 & 289 & 33,449 & 105\\
        \midrule
        LastFM & GMF  & 18.5 & 0.8 & 417.4 & 150 \\
        & MLP  & 1,657 & 71.4 & 37,257 & 162\\
        & NeuMF & 1,677 & 72.3 & 37,713 & 310\\
        \midrule
        Foursquare NY & GMF  & 13.5 & 1.4 & 114.2 & 461\\
        & MLP  & 1,200.2 & 128.5 & 10,195.9 & 473\\
        & NeuMF & 1,214.9 & 130.1 & 10,320.9 & 920\\
        \bottomrule
    \end{tabular}
    \caption{Computation and communication costs of the three considered models in each dataset.}
    \label{tab:flops_size}
\end{table*}
In this section, we evaluate the efficiency of the three considered models on the client side by measuring the Floating Point Operations (FLOPs) and the size of the parameters that need to be transferred to the coordination server for aggregation after local training. The number of FLOPs shows the computational overhead of the clients' devices for local training, while the size of the parameters shows the introduced communication overhead.

We report the results in Table \ref{tab:flops_size} for each of the four datasets considered. In the reported results, we show the average (Avg.), minimum (Min.) and maximum (Max.) FLOPs observed from clients in thousands, while the communication cost is measured in KB.

The computational demand in each model grows linearly with the number of local observations. For instance, the minimum number of observed instances in the MovieLens 100K and 1M datasets is 20. Note that during training, the observations grow with the negative feedback. Since both datasets contain a user with the same number of observations, the minimum FLOPs required are the same (3,200). The GMF model provides 89 and 90 times less computational cost on average than FedMLP and FedNeuMF, respectively.

The computational efficiency of GMF is also reflected in the size of the parameters that need to be transmitted for aggregation. The models' parameters grow linearly with the number of items in the item profile, while GMF has fewer communication requirements regarding the size of the parameters that are transferred for aggregation. More precisely, GMF requires the transmission of almost half as many parameters as the corresponding NeuMF model.

Comparing the GMF model with the traditional MF, the required number of FLOPs is almost equivalent. More specifically, the GMF model performs an element-wise multiplication of the user and an item's vector and the result is fed to a single processing unit, which is further transformed using the sigmoid function. The traditional MF model performs a dot product operation on the user and an item's vector. Hence, the additional computation cost of GMF compared to MF only concerns the final transformation of the sigmoid function for each item in the local profile. Similarly, the number of parameters to be transmitted for aggregation is almost equivalent. The additional parameters of the GMF model only concern the weights and the bias of the processing unit. In the considered setting, the GMF model only requires 13 additional float values, i.e. 0.1 KB, compared to traditional MF. 

Based on the observations on the computational cost, the communication cost and the offered recommendation quality of the three considered models, we argue that simple techniques are preferable to complex models due to their simplicity, at least in the federated setting, where mobile devices with low resources are involved. Compared to the hybrid NeuMF, the GMF model provides almost equivalent recommendations concerning the considered metrics and without incurring high computation and communication costs on the client side. Hence, NeuMF and complex models are preferred on devices with higher computational resources such as desktops, while simple models can be effectively trained on devices with low resources. Therefore, we identify a trade-off between quality and complexity in federated CF, i.e. adopting complex models in the federated setting is non-trivial as the overall computational cost may be prohibitive. In the future, we plan to evaluate the recommendation quality of other collaborative filtering algorithms and measure the trade-off between recommendations and computational cost.

\subsection{Heterogeneity impact}
In the federated setting, diversity concerning the computing resources and the size of local datasets is a common condition. In the considered datasets, quantity heterogeneity holds among clients, as observed in Table \ref{tab:data_hetero}. Hence, this section reports the computation time on clients concerning the system and data quantity heterogeneity. The additional computation cost on clients with limited resources or large training sets may lead to the straggling effect, which can heavily influence the global model's updates \cite{Tian2020FL}. 
\begin{figure*}[htb!]
    \centering
    \begin{subfigure}[b]{0.99\textwidth}
        \centering
        \includegraphics[width=0.85\textwidth]{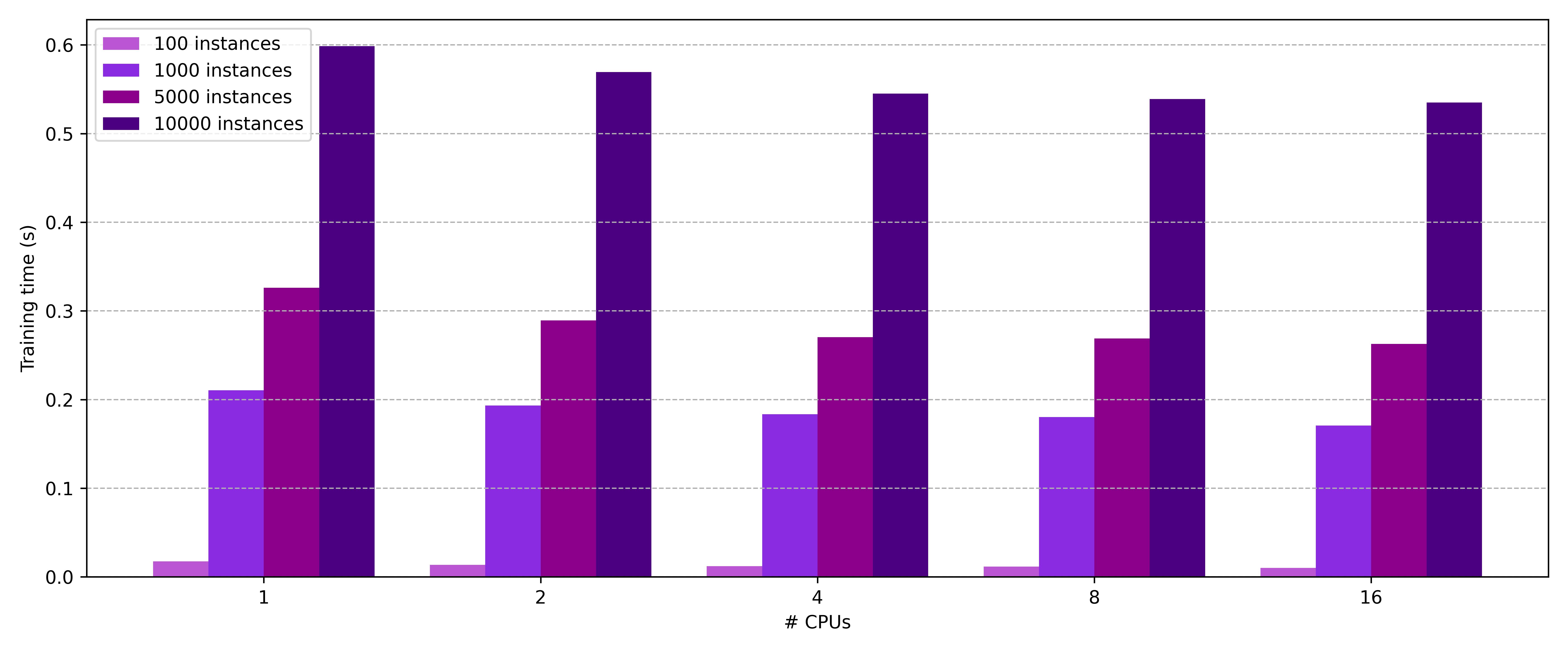}
        \caption{Training time per number of CPU resources and different number of training instances.}
    \end{subfigure}
    \begin{subfigure}[b]{0.99\textwidth}
        \centering
        \includegraphics[width=0.85\textwidth]{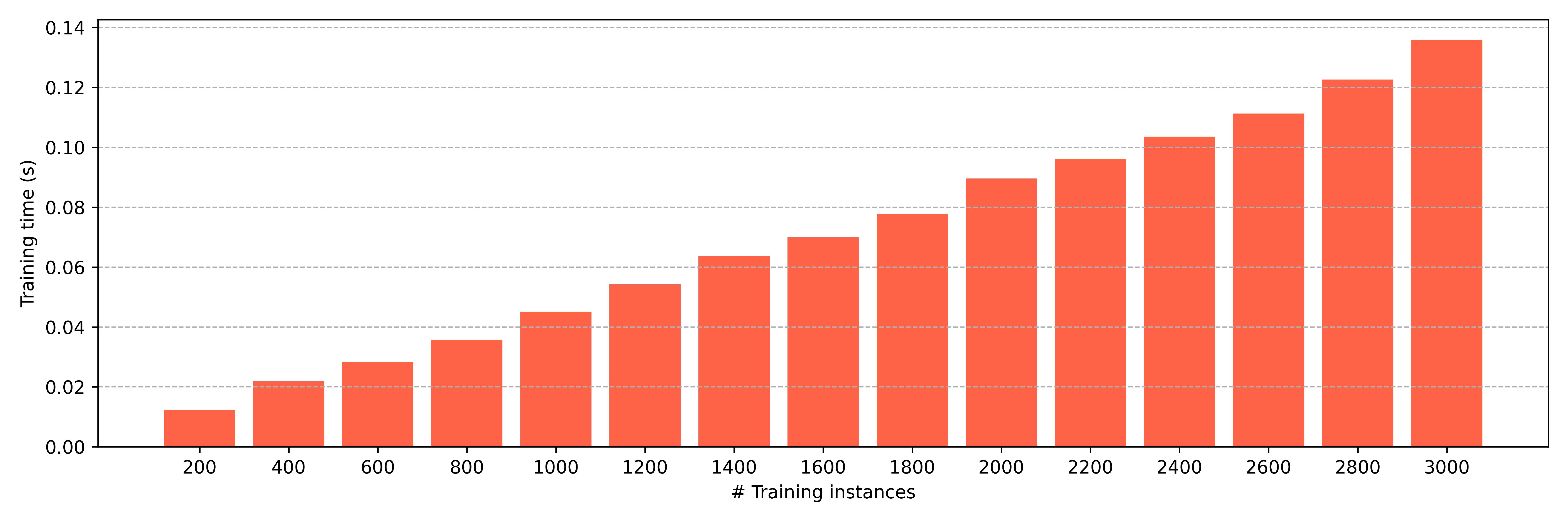}
        \caption{Training time with different number of training instances.}
    \end{subfigure}
\caption{Data quantity heterogeneity impact on the training time of the GMF model.}
\label{resource_data_hetero}
\end{figure*}

In Fig. \ref{resource_data_hetero}, we report the impact of resource and data quantity heterogeneity on the training time of the GMF model. We consider five cases of resources by considering \{1, 2, 4, 8, 16\} CPU cores and four cases regarding the size of the local dataset from \{100, 1000, 5000, 10000\} instances. In the worst case scenario, considering 10,000 observations with a single CPU, the training process is completed in less than a second, while using 16 CPU cores, the training is 10\% faster. In general, it is observed that the computation time gets longer with the increase of the size of the local dataset, while the system's heterogeneity does not heavily impact the computation time. Taking a closer look at the impact of the number of training instances, it is observed that the training time increases linearly with the number of local observations. For instance, the computation time with 2,000 samples is 0.089 $\approx$ 2 (0.045), where 0.045 is the training time using 5,000 instances. Based on these observations, the federated process may be heavily influenced by the amount of training data on each client side and should be further investigated to provide convergence guarantees for generalized federated learning.

\subsection{Number of participants impact}
In this section, we evaluate the robustness of both the centralized and federated settings concerning the number of users participating in the training process. We randomly selected a subset of users from $\{0.9, 0.8, 0.7, 0.5\}$ to participate in the computation and removed the rest of the users in each dataset. Note that we also excluded items that ended without interactions since cold-start problems are treated as a different task in CF \cite{Wei2017Cold}. We trained the GMF model from scratch by repeating the experiment 5 times by randomly selecting the users participating in the computation for each dataset.

\begin{figure*}[htb!]
    \centering
    \begin{subfigure}[b]{0.49\textwidth}
        \centering
        \includegraphics[width=0.92\textwidth]{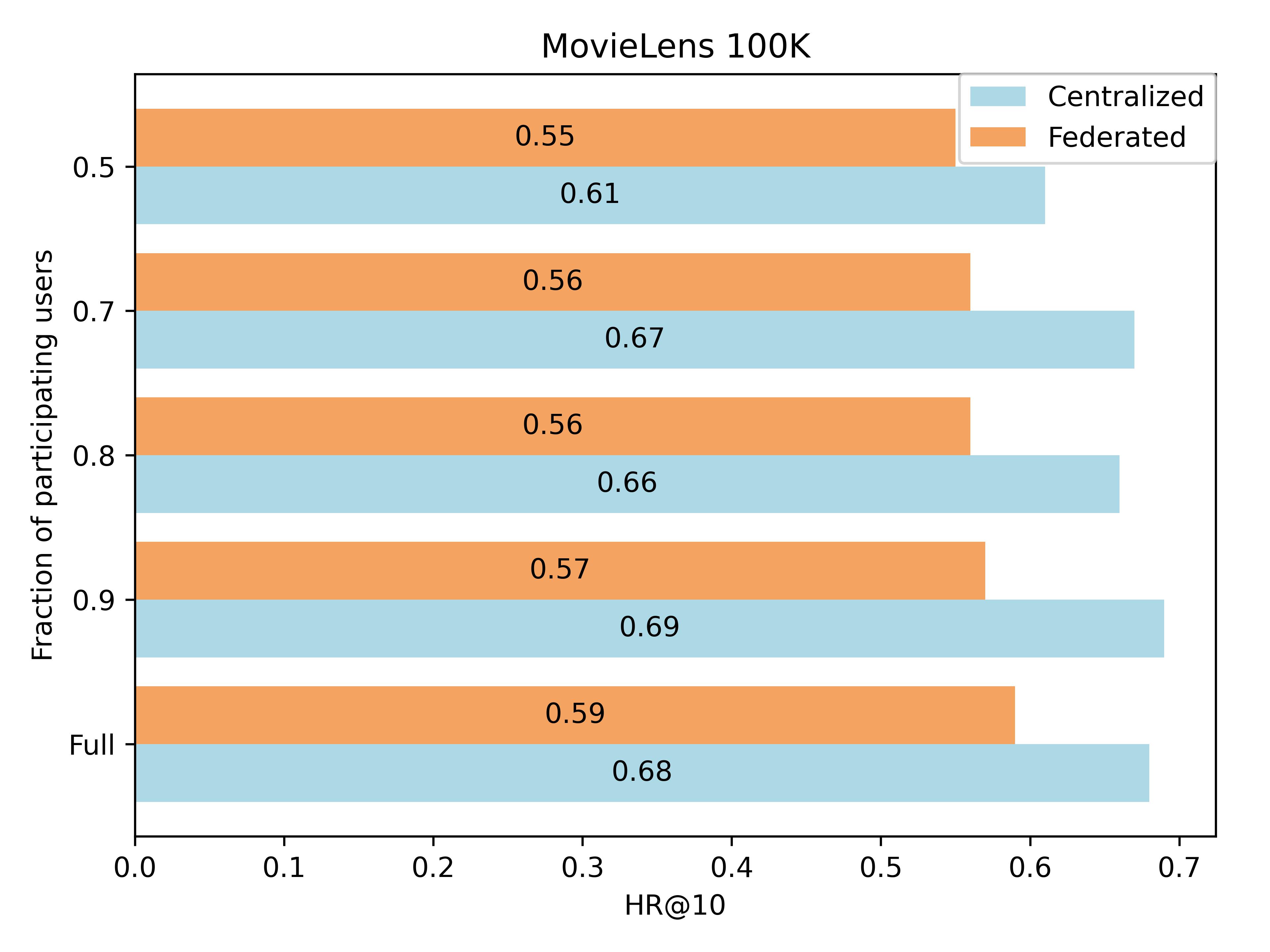}
    \end{subfigure}
    \begin{subfigure}[b]{0.49\textwidth}
        \centering
        \includegraphics[width=0.92\textwidth]{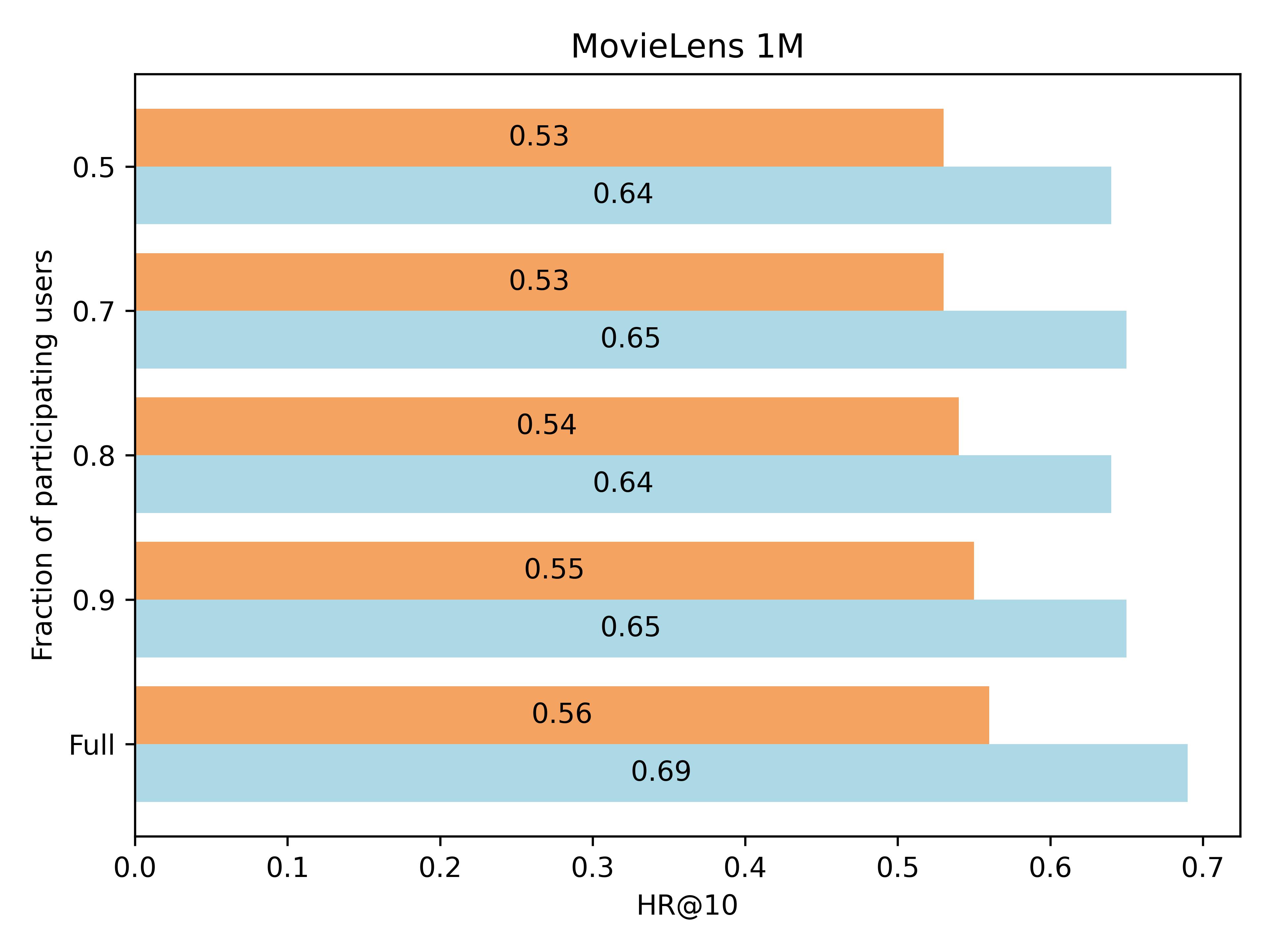}
    \end{subfigure}
    \begin{subfigure}[b]{0.49\textwidth}
        \centering
        \includegraphics[width=0.92\textwidth]{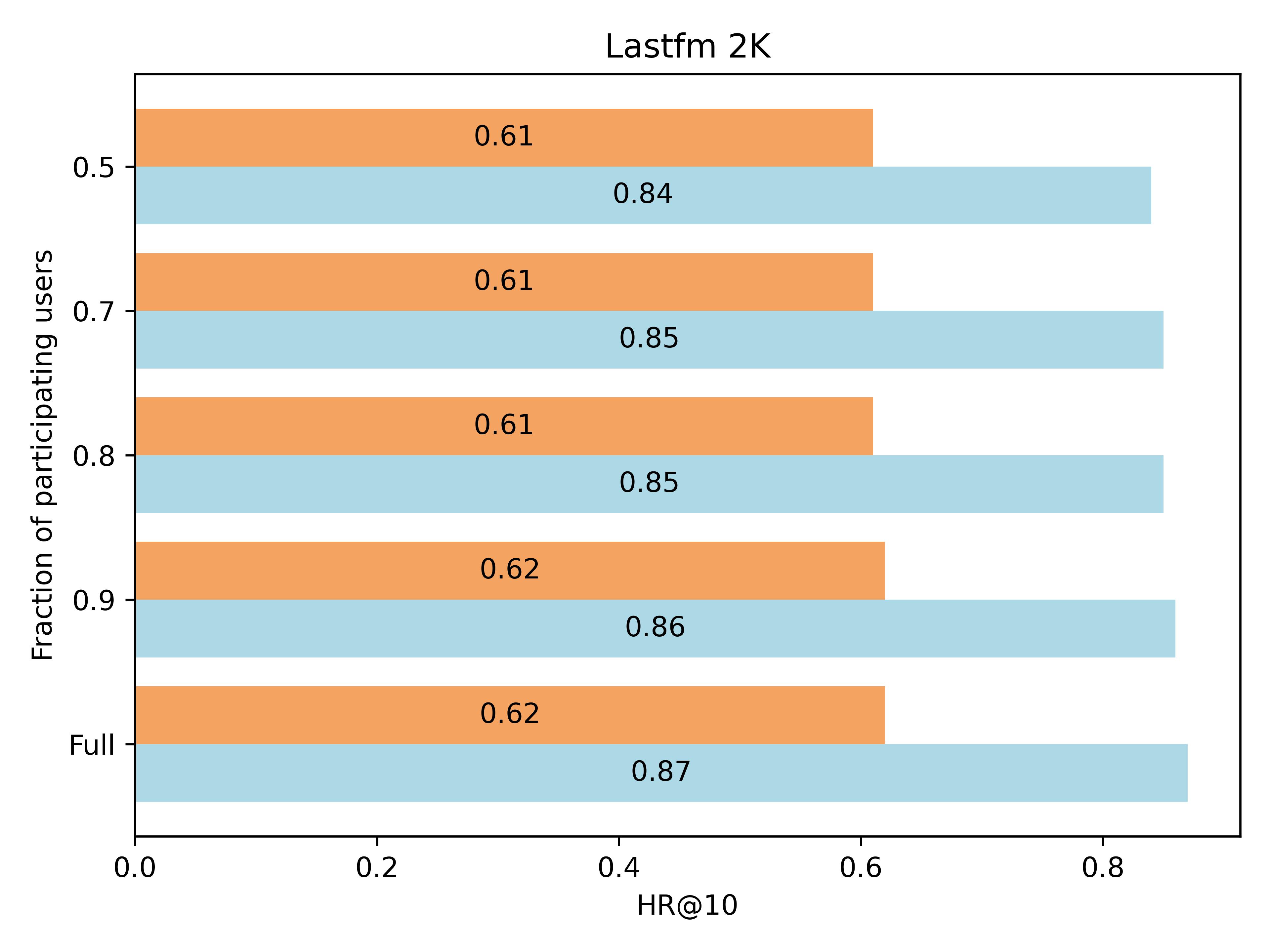}
    \end{subfigure}
    \begin{subfigure}[b]{0.49\textwidth}
        \centering
        \includegraphics[width=0.92\textwidth]{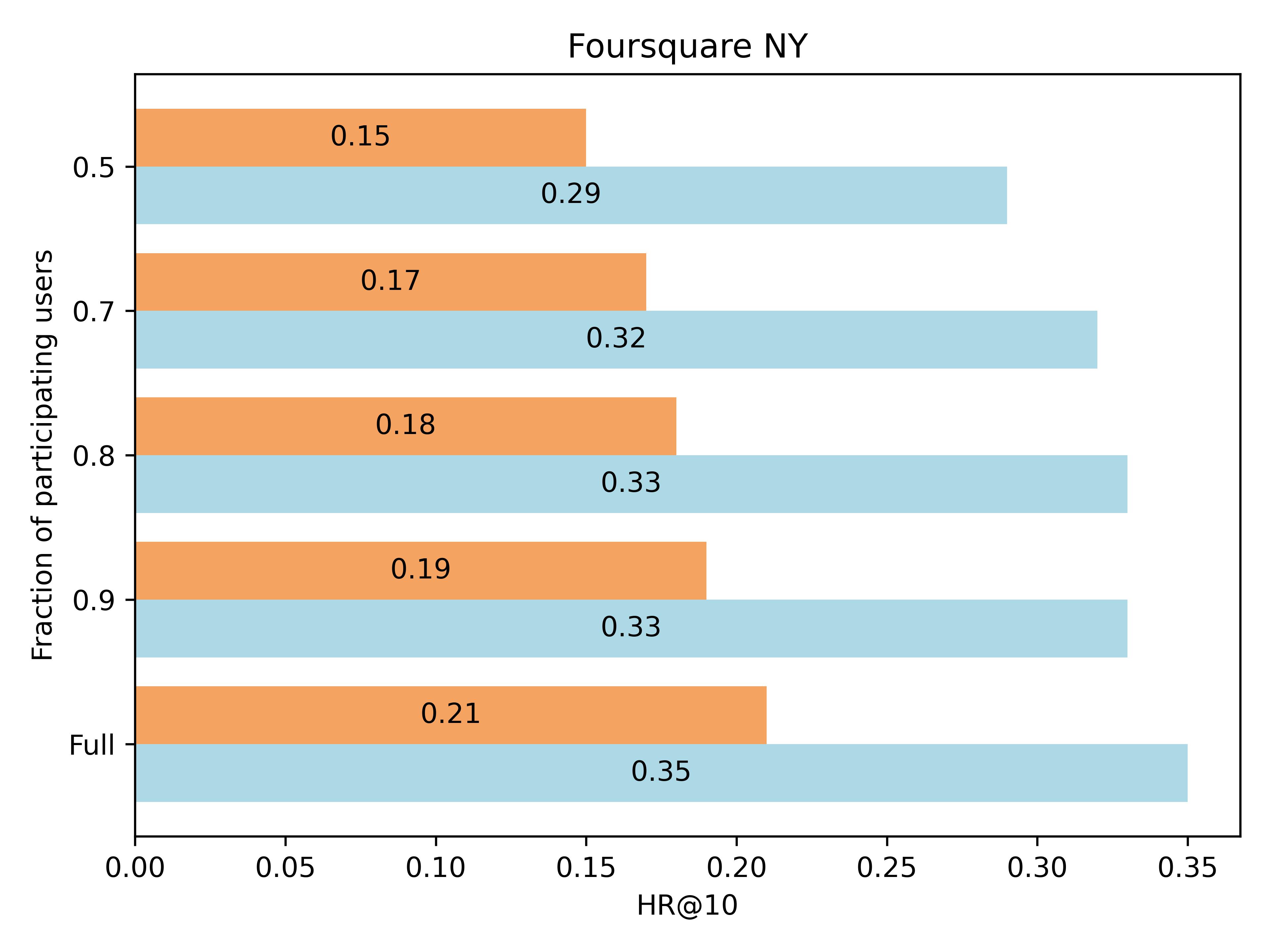}
    \end{subfigure}
\caption{Recommendation quality per fraction of participating users.}
\label{RobustFedNCF}
\end{figure*}
The averaged results are given in Fig. \ref{RobustFedNCF}. The recommendation quality by removing participating clients remains almost equivalent to the original settings. In both centralized and federated settings, there is a small decrease in the quality with the users' removal, between 3-7\% and 1-6\%, respectively. Hence, the finalized model in both techniques remains robust to the number of participants. More specifically, as the number of users increases, the quality of recommendations also increases. Considering that additional training iterations can further lead to better recommendations in the federated setting and the fact that this technique is applied to data that are not available in centralized environments, one can argue that FL has the potential to outperform traditional models in case of large-scale deployment.

\subsection{Secure aggregation impact}
The operations on randomly generated matrices with MF-SecAvg (Fig. \ref{MF-SecAvg_overview}, steps 2 and 3) will generate equivalent aggregated parameters to the plain aggregation approach. Therefore, the objective of this experiment is to gain insight into the computational overhead that is introduced. We choose to experiment with the FedGMF and FedNeuMF models because they comprise the smallest and largest model of the NCF approach, respectively. Table \ref{tab:SecAvgImpact} shows the average additional time required to initialize the random weights and generate the masked weights in the FedGMF and FedNeuMF models after a global round. Note that there is also an additional communication cost for the agreement phase of random seeds, but it is negligible, as mentioned in \cite{Bonawitz2017SecureAggregation}. 
\begin{table}[!ht]
  \centering
  \begin{tabular}{ lcc }
    \textbf{Dataset}    &   \textbf{FedGMF} & \textbf{FedNeuMF}  \\
    \midrule
    \textbf{MovieLens 100K}   & 3ms     &      7ms\\
    \textbf{MovieLens 1M}   &  5ms    & 12ms     \\
    \textbf{Lastfm 2K}   &   16ms   & 33ms     \\
    \textbf{Foursquare NY}  &  49ms    &  93ms   \\
    \bottomrule
  \end{tabular}
  \caption{MF-SecAvg impact on computation overhead in an aggregation round for each participant on a 4.0GHz, 8-core CPU.}
  \label{tab:SecAvgImpact}
\end{table}

The additional computational cost depends on the number of items included in each dataset. On the largest dataset (Foursquare NY), the overhead only concerns an additional computation time of 93ms with the NeuMF model. Thus, it is easily observed that the MF-SecAvg protocol has an imperceptible impact on the computation overhead. This suggests that integrating MF-SecAvg into federated recommender systems provides both high computational and communication efficiency while preserving the participants' privacy.

\section{Conclusion}
\label{Conclusion}
In this paper, we presented a federated version of the state-of-the-art method Neural Collaborative Filtering for generating high-quality recommendations. Although FL enables participants to build independent models without revealing their raw data, private interactions can still be leaked to the coordination through the output of their computations. To overcome this challenge, we presented a privacy-preserving approach by adapting the SecAvg protocol \cite{Bonawitz2017SecureAggregation} to meet the update process of latent factor models. We evaluated the recommendation quality and the efficiency of FedNCF and discussed the impact of the utilized aggregation function. Our experiments demonstrated the feasibility of FL in recommender systems and supported the concept that small contributions from low-resource computing nodes whose data remain local can lead to a high-quality machine learning model.

A critical future direction for improving FL is further focusing on its security and privacy analysis. Even though FL enables a higher privacy level than centralized learning due to data locality, a formal estimation of the information leakage is crucial. While secure aggregation protocols such as \cite{Bonawitz2017SecureAggregation}, ensure the privacy of the participants in a single round, the privacy guarantee across multiple rounds is still an open issue.

\section*{Acknowledgments}
This work was supported by the European Union and Greek national funds through the Operational Program Competitiveness, Entrepreneurship and Innovation, under the call RESEARCH - CREATE - INNOVATE (project code:  T1EDK-02474, grant no. MIS 5030446).

 \bibliographystyle{elsarticle-num} 
 \bibliography{cas-refs}





\end{document}